\documentclass[12pt]{article}
\usepackage{amsmath, amssymb}

\evensidemargin \oddsidemargin
\newcommand{\mybaselineskip}{\baselineskip 20pt}

\newtheorem{theorem}{Theorem}[section]          

\newtheorem{lemma}[theorem]{Lemma}
\newtheorem{proposition}[theorem]{Proposition}
\numberwithin{equation}{section}                
\renewcommand{\theequation}{\arabic{section}.\arabic{equation}}

\newcommand{\R}{\mathbb{R}}
\newcommand{\Complex}{\mathbb{C}}
\renewcommand{\Re}{\mathop{\mathrm{Re}}}
\renewcommand{\Im}{\mathop{\mathrm{Im}}}

\newcommand{\eigen}{\mathtt{E}_1}

\newcommand{\myspan}{\mathop{\rm span}}

\newcommand{\longto}{\longrightarrow}
\newcommand{\norm}[1]{\left\Vert #1 \right\Vert}
\newcommand{\length}[1]{\left| #1 \right|}
\newcommand{\bkA}[1]{\left \langle #1 \right \rangle}
\newcommand{\bke}[1]{\left( #1 \right)}
\newcommand{\bkt}[1]{\left[ #1 \right]}
\newcommand{\bket}[1]{\left\{ #1 \right\}}

\newcommand{\e}{\varepsilon}

\newcommand{\loc}{_{\mathrm{loc}}}

\newcommand{\myremark}{\addtocounter{equation}{1}{{\bf Remark
(\theequation)}}\quad}

\newcommand{\myproof}{\noindent {\bf Proof.}\quad}
\newcommand{\myendproof}{\hspace*{\fill}{{\bf \small Q.E.D.}} \vspace{10pt}}

\newcommand{\al}{\alpha}

\renewcommand{\L}{\mathcal{L}}
\newcommand{\la}{\lambda}

\newcommand{\Pc}{\, \mathbf{P}\! _\mathrm{c} \, \! }

\newcommand{\PcH}{\, \mathbf{P}\! _\mathrm{c} \! ^{H_0} \,}
\newcommand{\PcL}{\, \mathbf{P}\! _\mathrm{c} \! ^\L \,}
\newcommand{\Hc}{\, \mathtt{H} _\mathrm{c} \, \! }

\newcommand{\wt}{\widetilde}
\newcommand{\conj}{\mathtt{C}}

\newcommand{\pd}{\partial}
\newcommand{\donothing}[1]{}

\newcommand{\wbar}[1]{\overline{\rule{0pt}{2.4mm} {#1}}}
\newcommand{\lbar}[1]{\underline{#1}}
\newcommand{\n}{{n}}  

\newcommand{\leC}{\,\lesssim\,}  

\newcommand{\xm}{|x|}

\newcommand{\xitwo}{\xi^{(2)}}
\newcommand{\xithree}{\xi^{(3)}}  
\newcommand{\uu}{\mu} 
\newcommand{\vv}{\nu} 

\begin{document}
\mybaselineskip

\title{Relaxation of Excited States in Nonlinear Schr\"odinger Equations}

\author{Tai-Peng Tsai\footnote{ttsai@cims.nyu.edu} \qquad
Horng-Tzer Yau\footnote{Work partially supported by NSF grant
DMS-0072098,  yau@cims.nyu.edu} \\ \vspace*{-0.3cm} \\ Courant
Institute, New York University} 

\date{August 27, 2001}


\donothing{
\author{Tai-Peng Tsai}
\address{Courant Institute, New York University, New York, NY 10012, USA}
\email{ttsai@cims.nyu.edu}

\author{Horng-Tzer Yau}
\address{Courant Institute, New York University, New York, NY 10012, USA}
\email{yau@cims.nyu.edu}
\thanks{The work of the second author was partially supported by 
NSF grant DMS-0072098}


\date{January 27, 2001}

}

\maketitle


\begin{abstract}

We consider a nonlinear Schr\"odinger equation in $\R^3$ with a bounded
local potential. The linear Hamiltonian is assumed to have two bound states with the
eigenvalues satisfying some resonance condition. Suppose that the
initial data is small and is near some nonlinear
\!{\it excited}\, state.  We give a sufficient condition
on the initial data so that
the solution to the nonlinear Schr\"odinger equation
approaches to certain
nonlinear {\it ground} state as the time tends to infinity.
\end{abstract}



\section{Introduction}

Consider the nonlinear  Schr\"odinger equation
\begin{equation} \label{Sch}
i \pd _t \psi = (-\Delta + V) \psi + \la |\psi|^2 \psi, \qquad
\psi(t=0)= \psi_0
\end{equation}
where  $V$ is a smooth localized potential,  $\la $ is an order $1$
parameter and $\psi=\psi(t,x):\R\times \R^3 \longto \Complex$  is a
wave function. The goal of this paper is to study the asymptotic
dynamics of the solution for  initial data $\psi_0$ near some
{\it nonlinear excited state}.

Recall that  for any solution $\psi(t)\in H^1(\R^3)$ the $L^2$-norm
and the Hamiltonian
\begin{equation} \label{1-2}
{\mathcal H}[\psi] = \int \frac 12 |\nabla \psi|^2 +  \frac 12 V |\psi|^2
+ \frac 14 \la |\psi|^4 \, d x ~,
\end{equation}
are constants  for all $t$. The global well-posedness for small solutions
in $H^1(\R^3)$ can be proved using these conserved quantities and a
continuity argument.

We assume that the linear Hamiltonian $H_0 :=- \Delta + V$
has two simple eigenvalues $e_0<e_1<0$
with normalized eigen-functions $\phi_0$, $\phi_1$. We further assume
that
\begin{equation}   \label{evcon}
  e_0 < 2 \, e_1 ~.
\end{equation}
The nonlinear bound states to the Schr\"odinger equation \eqref{Sch}
are solutions to the equation
\begin{equation}   \label{Q.eq}
    (-\Delta + V) Q + \la |Q|^2 Q = EQ  ~.
\end{equation}
They are critical points to the Hamiltonian ${\mathcal H}[\phi] $
defined in \eqref{1-2}
subject to the constraint that the $L^2$-norm of $\psi$ is fixed.
For any bound state
$Q=Q_E$, $\psi(t) = Q e^{-i E t } $ is a solution to the nonlinear
Schr\"odinger equation.

We may obtain two families of such bound
states by standard bifurcation theory, corresponding to the two
eigenvalues of  the linear Hamiltonian. For any $E$ sufficiently
close to $e_0$ so that $E-e_0$ and $\la$
have the same sign, there is a unique  positive solution $Q=Q_E$ to
\eqref{Q.eq} which decays exponentially  as $x \to \infty$. See
Lemma \ref{th:2-1}. We call this family the {\it nonlinear ground
states} and we refer to it as $\bket{Q_{E}}_{E}$. Similarly,
there is a {\it nonlinear excited state} family $\bket{Q_{1,E}}_E$. We
will abbreviate them as $Q$ and $Q_{1}$. From Lemma \ref{th:2-1},
we also have $\norm{Q_{E}} \sim
|E-e_0|^{1/2}$ and $\norm{Q_{1,E}} \sim |E-e_1|^{1/2}$.

It is well-known that the
family of nonlinear ground states is stable in the sense that if
\[
   \inf_{\Theta, E} \norm{ \psi(t) -Q_E \, e^{i \Theta} }_{L^2}
\]
is small for $t=0$, it remains so for all $t$, see \cite{RW}.
Let $\norm{ \cdot
}_{L^2 \loc}$ denote a local $L^2$ norm ( a precise choice will be
made later on).  One expects that this difference actually
approaches  zero in local $L^2$ norm, i.e.,
\begin{equation}\label{as}
\lim_{t \to \infty} \inf_{ \Theta, E} \norm{ \psi(t) -Q_E \, e^{i
\Theta} }_{L^2 \loc} = 0 ~.
\end{equation}
If $- \Delta + V$ has only one bound state, it is proved in
\cite{SW1} \cite{PW} that the evolution will eventually settle down to some
ground state $Q_{E_\infty}$ with $E_\infty$ close to $E$. Suppose now
that  $- \Delta + V$ has multiple bound states, say, two bound
states: a ground state $\phi_0$ with eigenvalue $e_0$ and an excited
state $\phi_1$ with eigenvalue $e_1$.
It is proved in \cite{TY} that the
evolution with initial data $\psi_0$ near some $Q_E$ will eventually
settle down to some ground state $Q_{E_\infty}$ with $E_\infty$ close
to $E$.  See also \cite{BP} for the one dimensional case
and \cite{SW2} for nonlinear Klain-Gorden equations.

Denote by  $L^2_{r}$ the weighted $L^2$ spaces ($r$ may be
positive or negative)
\begin{equation} \label{L2r.def}
L^2_{r}\,(\R^3) \; \equiv \; \bket{\phi \in L^2(\R^3) \; : \;
\bkA{x}^r \phi \in L^2(\R^3) } ~.
\end{equation}
The space for initial data in \cite{TY} is
\begin{equation} \label{Y.def}
Y \equiv H^1(\R^3) \cap L^2_{r_0}\,(\R^3) ~,
 \qquad r_0>3 ~.
\end{equation}
We shall use $L^2\loc$ to denote $L^2 _{- r_0}$.
The parameter $r_0>3$ is fixed and we can choose,
 say, $r_0=4$ for the rest of this paper.
We now state the assumptions in \cite{TY}
on the potential $V$.

\noindent {\bf Assumption A0}: $- \Delta + V$ acting on $L^2(\R^3)$
has $2$ simple eigenvalues $e_0<e_1< 0 $, with normalized
eigenvectors $\phi_0$ and $\phi_1$.

\noindent {\bf Assumption A1}: Resonance condition. Let $e_{01}=
e_1-e_0$ be the spectral gap of the ground state. We assume that
$2e_{01} > |e_0|$ so that $2e_{01}$ is in the continuum spectrum of
$H_1$. Let
\begin{equation}
\label{gamma0.def}
\gamma_0 := \lim_{\sigma \to 0+} \Im \bke{\phi_0 \phi_1^2, \,
\frac 1 {H_0 + e_0 - 2 e_1 -\sigma i} \Pc^{H_0}\phi_0 \phi_1^2}
\end{equation}
Since the expression is quadratic, we have $\gamma_0 \ge 0$.
We assume, for some $s_0< C n_0^2$ small enough,
\begin{equation}
\label{A:gamma0}
\inf_{|s|<s_0} \lim_{\sigma \to 0+} \Im \bke{\phi_0 \phi_1^2, \,
\frac 1 {H_0 + e_0 - 2 e_1 +s -\sigma i} \Pc^{H_0}\phi_0 \phi_1^2} \ge
\frac 34 \gamma_0 > 0 ~.
\end{equation}
We shall use $0i$ to replace  $\sigma i$ and the limit
$\lim_{\sigma \to 0+}$ later on.

\noindent {\bf Assumption A2}: For $\la Q_E^2$ sufficiently small,
the bottom of the continuous spectrum to $-\Delta+ V+\la Q_E^2$,
$0$, is not a generalized eigenvalue, i.e., not a resonance. Also,
we assume that $V$ satisfies the assumption in  \cite{Y} so that
the $W^{k,p}$ estimates  $k\le 2$ for the wave operator
$W_H=\lim_{t\to \infty} e^{i t H}e^{it(\Delta+E)}$ hold for $k \le
2$, i.e., there is  a small $\sigma>0$ such that,
\[
|\nabla^\al V(x)| \le C \bkA{x}^{-5-\sigma}, \qquad
\text{for } |\al|\le 2 ~.
\]
Also, the functions $(x\cdot \nabla)^k V$, for $k=0,1,2,3$, are
$-\Delta$ bounded with a $-\Delta$-bound $<1$:
\begin{equation*}
  \norm{(x\cdot \nabla)^k V\phi}_2 \le \sigma_0 \norm{-\Delta\phi}_2 +
C\norm{\phi}_2,
  \qquad \sigma_0 < 1 , \quad k=0,1,2,3  ~.
\end{equation*}

\bigskip

Assumption A2 contains some standard conditions to assure that most
tools in linear Schr\"odinger operators apply. These conditions are
certainly not optimal. The main assumption in A0-A2 is the condition
$2e_{01} > |e_0|$ in
assumption A1. The rest of assumption A1 are just generic assumptions.
This condition states that the excited state energy is closer
to the continuum spectrum than to the ground state energy. It
guarantees that twice the excited state energy of $H_1$
(which one obtains from taking the square of the excited state
component) becomes a resonance in the continuum spectrum
(of $H_1$). This resonance produces the main relaxation mechanism.
If this condition fails, the resonance occurs in higher order terms
and a proof of relaxation will be much more complicated. Also, the rate
of decay will be different.

The main result in  \cite{TY}
concerning the relaxation of the ground states can be summarized
in  the following
theorem.

\medskip
\noindent {\bf Theorem A} {\it Suppose  that suitable assumptions on
$V$ hold. Then there are small universal constants $\e_0, n_0>0$ such
that,  if the initial data $\psi_0$ satisfies  $\norm{\psi_0-\n
e^{i\Theta_0}\phi_0}_Y \le \e_0^{2} \n^2$ for some $\n\le \n_0$ and
some $\Theta_0\in \R$,  then there exists an $E_\infty$  and a
function $\Theta(t)$ such that $\norm{Q_{E_\infty}}_Y - \n=O(\e_0^{2}
\n)$, $\Theta(t) = -E_\infty t+ O(\log t)$ and}
\begin{equation}
\norm{\psi(t)-Q_{E_\infty} e^{ i \Theta(t)}}_{L^2 \loc} \le
C(1+t)^{-1/2}  ~.
\end{equation}

\medskip

This theorem settles the question of asymptotic profile near
ground states.
Suppose that the initial data $\psi_0$ is now near some
nonlinear excited state. From the physical ground, we expect that
$\psi_t$  will eventually decay to some ground state unless
the initial data $\psi_0$ is exactly a nonlinear  excited state.
We call this the "strong relaxation property". For comparison,
we define a weaker property, the "generic relaxation property",
as follows.
Denote  the space of initial data by $X$. Let $X_1$ ($X_0$ resp.) be
the subspace of initial data such that the asymptotic profiles
are given by some nonlinear excited (ground resp.) states.
We shall say that
the dynamics satisfy the generic relaxation property if
$X_1$ has "measure zero".
This concept depends on a notion of measure
which should be specified in each context.

With this
definition, the strong relaxation property means that $X_1$
is exactly the space of excited states. In particular,
$X_1$ is finite dimensional. We first note that
the strong relaxation property is
false. For any nonlinear excited state $Q_1$, define
$X_{1, Q_1}$ to be the set of initial data converging to
$Q_1$ asymptotically.
It is proved in \cite{TY2} that for any given
nonlinear excited state $Q_1$,  $X_{1, Q_1}$ contains
a  finite co-dimensional set.
Thus our goal is to establish
some weaker statement such as
the generic relaxation property. This is the
first step toward a classification of  asymptotic
dynamics of the nonlinear Schr\"dinger equation.

\donothing{
We can visualize these results in the following picture.
Since the
excited states and the ground states  depend
on  the two parameters, the masses and the phases,
they are two dimensional tubes. Each circle in
these two  tubes can be viewed as a  fixed "point" of the dynamics
in the sense that the evolution is just a rotation of the circle.
On the other hand, each circle
by itself is not stable since a small perturbation
will in general change the mass. The
family of the ground states taken as a whole is however stable
under perturbation  \cite{TY}.
The family of excited states,
even taken as a whole, is not stable, due to that
$X_{1, Q_1}$ contains
a  finite co-dimensional set \cite{TY2}.
}

In this paper, we shall prove that for any  excited state,
there is a small neighborhood $\mathcal N$ so that
$$
|{\mathcal N} \cap X_1 | \le C \|\psi_0\|^2 |{\mathcal N} \cap X_0 |
$$
This estimates states that the ratio
between  $X_1$  and $X$  around excited states are
bounded by the mass of the initial  wave function $ \|\psi_0\|^2$.
(Since we have not given a measure on the space of initial data,
this statement is not well-defined and a precise statement will be given
later on.)

\donothing{
 Recall our aim is to prove that $X_1$
has measure zero. Our result  have shows only  that $X_1$ is small
relative to $X$ when the mass of the initial data is small.
However, we have explicit estimates on the set $X\setminus
X_1$.
}

In order to state the main result, we
first decompose the wave function using
the eigenspaces of  the Hamiltonian  $H_0$ as
\begin{equation} \label{psidec0} \psi = \lbar{x}\phi_0 + \lbar{y}\phi_1
+\lbar{\xi} , \quad \lbar{\xi} = \Pc^{H_0}\, \psi ~. \end{equation}
For initial data near excited states, this decomposition
contains an error of order $y_0^3$ and it is difficult to
read from \eqref{psidec0} whether the wave function is exactly
an excited state.  Thus  we shall use the decomposition
\begin{equation} \label{psidec1} \psi = {x}\phi_0 + Q_1(y) +{\xi} ~. 
\end{equation}
where
\begin{equation} y=\lbar{y}, \quad x =\lbar{x} -(\phi_0, Q_1(y)) ,\quad
\xi=\lbar{\xi}-\Pc Q_1(y) ~.
 \end{equation}
Here we have used the convention that
$$
Q_1(y) :=Q_1(m) e^{i \Theta}, \qquad m= |y|, \; \;
m e^{i \Theta} = y
$$
%
%
%
We shall prove that for $\psi$ with sufficiently small $Y$ norm
\eqref{Y.def},
such a decomposition exists and is unique in section 2.
Thus we assume that  $\psi_0= {x_0}\phi_0 + Q_1(y_0) +{\xi_0}$ is
sufficiently small in $Y$. Let $n_0$ and $\e_0$ be the same small
constants given  in Theorem A. By choosing a smaller $n_0$, we may assume
$n_0 \le \e_0^2/4$. We assume the initial data satisfies that
\begin{equation} \label{eq:1-10}
\begin{split}
&\norm{\psi_0}_{Y} =\n, \ 0 < n \le n_0, \qquad |y_0| \ge  \tfrac 12 \n , \\
&|x_0| \ge 2 \n \, e^{-1/n} , \qquad |x_0| \ge  \e_2^{-1} n^2
\norm{\xi_0}_Y ~,
\end{split}
\end{equation}
where $\e_2>0$ is a small universal constant to be fixed later in the proof.
These conditions can be interpreted as follows: The excited
state component, $y_0$, should account for at least half the mass
of the initial data. Under this condition, if the ground state
component, $x_0$ is not too small compared with the continuum
component $\xi$, then the dynamics relaxes to some ground state.
The
condition $|x_0| \ge 2 \n \, e^{-1/n}$ is a very mild assumption
to make sure that $x_0$ is not incredibly small.
The following constant  will be used in many places in this paper.
Define
\begin{equation}\label{e.def}
\e:= \min \bket{ \e_0/2, \, ( \log (2n/x_0))^{-1/2}}~.
\end{equation}
Since $ n \le n_0\le \e_0^2/4$ and $|x_0| \ge  2 \n \, e^{-1/n}$, we
have $n \le \e^2$.

\begin{theorem}  \label{th:1-1}
Suppose  the assumptions on $V$ given above hold.  Let $\psi(t,x)$ be
a solution of \eqref{Sch} with the initial data $\psi_0$ satisfying
\eqref{eq:1-10}. Let
\begin{equation}\label{n1.def}
n_1 = \bke{|x_0|^2 + \tfrac 12 \,|y_0|^2 }^{1/2}
\sim n~.
\end{equation}
Then, there exists an $E_\infty$ and a function $\Theta(t)$ such
that $\norm{Q_{E_\infty}}_Y - n_1=O(\e \n)$, $\Theta(t) =
-E_\infty t+ O(\log t)$ and
\begin{equation}
C_1(1+t)^{-1/2}  \le \norm{\psi(t)-Q_{E_\infty} e^{ i
\Theta(t)}}_{L^2 \loc} \le C_2(1+t)^{-1/2}  ~.
\end{equation}
for some constants $C_1$ and $C_2$.
\end{theorem}


It is instructive to  compare our result with the linear stability
analysis of \cite{S}  \cite{SS}   \cite{G} \cite{G2} and
\cite{SS2}. In our setup the main result in \cite{G2} states that
the linearized operator around a nonlinear excited state is
structural stable if $e_0 < 2 \, e_1$ and unstable if
$e_0 > 2 \, e_1$. Hence the excited states considered in this article is
unstable and is expected to decay under generic perturbations. The
instability of the excited state contains in  Theorem 1.1 is thus
consistent with the linear analysis. Notice that Theorem 1.1 tracks the
dynamics for all time including time regime when the dynamics are
far away from the excited states. Furthermore,
{\it for all initial data considered in Theorem 1.1,
the  relaxation rate to the asymptotic ground state
is exactly of order $t^{-1/2}$}, a rate very different from the standard
linear Schr\"odinger equations.

In view of the linear analysis,
the existence \cite{TY2} of (nonlinear) stable directions for
excited states
is a more surprising result.
For the linear stable case,  i.e., $e_0 > 2 \, e_1$,
the only  rigorous result is the existence \cite{TY2}
of (nonlinear) stable directions in this case. Although
the linear analysis states that all directions are linearly
stable, on physics ground we still expect
excited states remains generically  unstable.

We now explain the main idea of the proof for Theorem 1.1.
The relaxation mechanism
can be divided into three time regimes:

1. {\bf The initial layer}: The component of the wave function
in  the continuum spectrum direction
gradually disperses away;
the  components in the bound states directions  do not change much.

2. {\bf The transition regime}:
Transition from the  excited state to the ground state takes
place in this interval. The component along the
ground state
grows in this regime; that  along the excited state
is slightly more complicated. We can further
divide this time regime into  two intervals. In
part (i), the component along the excited state
does not change much. In
part (ii), it decreases steadily and eventually becomes smaller
than the component along the
ground state.

3. {\bf Stabilization}:
The ground state dominates and is stable. Both the excited states
and dispersive part gradually decay.

In different time regimes, the dominant terms are different
and we have to linearize the dynamics according to the
dominant terms. In the first time region,
$\psi(t)$ is near an excited state,
and it is best to use
operator linearized around the excited state.
In the third time regimes,
$\psi(t)$ is near a ground states, and it is best to use
operator linearized around  a ground state.
For the transition regimes, the dynamics are far away from
both excited and ground states and we can only use
the linear Hamiltonian  $H_0$.

Besides technical problems associated with
changing  coordinate systems
in different time intervals, there is an intrinsic difficulty
related to the time reversibility of the Schrodinger equation.
Imagine that  we are now ready to
show that our dynamics is in the third time regime and will
stabilize around some nonlinear ground state. If we take the
wave function $\psi_t$ at this time and time reverse the dynamics,
then the dynamics will drive this wave function back to the
initial state near some excited state. The time reversed state
  $\psi_t$ and  the wave
function $\psi_t$ itself will satisfy the same  estimates in the usual
Sobolev or $L_p$ senses. However, their dynamics are completely
different: one stabilizes to a ground  state; the other back to
near an excited state. This suggests that $\psi_t$ carries information
concerning the time direction and this information will not show
if we measure it by the usual estimates.

This time reversal difficulty manifest itself in the technical proofs as
follows. We shall see that,  when the third
time regime begin, the  dispersive
part is not well-localized and its
$L^2$-norm   can be larger
than that of the bound states---both violate conditions
for approaching to ground states in \cite {TY}.
To resolve this issue, we need to extract information which are
time-direction sensitive so that even though the disperive part
may be large, it is irrelevant since it is ``out-going''.
Though the concept of ``out-going'' wave is known for  linear
Schr\"odinger equations, it is difficult to implement it for
nonlinear Schr\"odinger equations. Our strategy is to identify
the main terms of the dispersive part and
 calculate them   explicitly. These terms carry
sufficient information concerning the time direction.
The rest are  error terms
and we can use various Sobolev or $L_p$ estimates.

{\bf Resonance induced decay and growth}

To illustrate the mechanism of resonance induced decay and growth,
we consider the problem in the coordinates
with respect to  the linear Hamiltonian $H_0 = - \Delta + V$,
\[
\psi(t) = {x(t)}\phi_0 + y(t) \phi_1 +{\xi(t)} ~, \quad, \xi(t)
\in  \PcH\psi (t)
\]
The nonlinear term  $\psi^2 \bar \psi$, (assume $\la=1$)
can be split into a sum of many terms using this
decomposition. However, we claim that there is only one
important nonlinear term in the equation for each component:
\begin{eqnarray}
i\dot x &=& e_0 x + (\phi_0, (y \phi_1)^2 \bar \xi) + \cdots
\label{model1}
\\
i\dot y &=& e_1  y + (\phi_1, 2(x \phi_0)(\bar y \phi_1) \xi)+ \cdots
\label{model2}
\\
i\pd_t \xi &=& H_0 \xi + \PcH \bar x y^2 \phi_0\phi_1^2+ \cdots
\label{model3}
\end{eqnarray}
From \eqref{model1}, we know $ u(t) = e^{i e_0 t} x(t)$
has less oscillation of lower order than $x(t)$.
Hence we say $x(t)$ has a  phase factor
$-e_0$. Similarly,  $y(t)$ has a phase factor $-e_1$. The nonlinear term
$\bar x y^2 \phi_0\phi_1^2$ has a phase factor $e_0 - 2e_1$, which,
due to the assumption \eqref{evcon},  is
the only term in $\psi^2 \bar \psi$ with a negative phase factor.
It gives a term in $\xi$:
\[
\xi(t) = \bar x y^2(t) \Phi + \cdots, \qquad
\Phi=\frac 1{H +e_0 - 2e_1-0i} \, \phi_0\phi_1^2
\]
Notice that $\Phi$ is complex.
Substituting this term into \eqref{model1} and \eqref{model2}, we
have
\begin{equation} \label{model4}
\begin{aligned}
i \dot x &= i \gamma_0 |y|^4 x  + \cdots,
\\
i \dot y &= -2 i \gamma_0 |x|^2 |y|^2 y + \cdots,
\end{aligned}
\end{equation}
with $\gamma_0$ given in \eqref{gamma0.def}.
In \eqref{model4} we have omitted two types of irrelevant terms:

1. Terms with same phase factors as $x$ or $y$:
for example, $e_0 x$ and $|y|^2 x$ in \eqref{model1}.
Since their coefficients are real, they disappear when we consider the
equations for $|x|$ and $|y|$.

2. Terms with different phase factors: for example, $\bar x y^2$
in \eqref{model1}.
Since these terms have different phases, their contribution
averaging over time will be small.
This can be made precise by the  Poincar\'e normal form.

From \eqref{model4} we obtain  the decay of $y$ and the growth of $x$
as well as the three time regimes mentioned previously.
However, it should be warned that
this set-up is only suitable when both $x$ and $y$ are
of similar sizes.

\section{The initial layer and the transition regimes: The set up}

We now  outline the basic strategy for
the initial layer and the transition regimes. We first review the
construction of the bound state families.

\subsection{Nonlinear bound states}

The basic properties of the ground state families
can be summarized in the following lemma from  \cite{TY}.

\begin{lemma} \label{th:2-1}
Suppose that $-\Delta+V$ satisfies the assumptions (A0) and (A2).
Then there is an $\n_0$ sufficiently small such that for $E$ between
$e_0$ and $e_0+ \la \n_0^2$  there is a nonlinear ground states
$\bket{Q_{E}}_E$ solving  \eqref{Q.eq}. The nonlinear ground state
$Q_{E}$ is real, local, smooth, $ \la^{-1}(E-e_0)>0$, and
\[
Q_{E} = \n \,\phi_0 +O(\n ^{3 }) ~,\qquad \n \approx C \,
[\la^{-1}(E-e_0)]^{1/2} \; , \quad C=(\int \phi_0^4 \, dx)^{-1/2}.
\]
Moreover, we have $R_E=\pd_E Q_{E} = O(\n^{-2} ) \, Q_{E} + O(\n  )=
O(\n^{-1} )$ and $\pd_E^2 Q_{E} = O(\n^{-3} )$. If we define
$c_1\equiv (Q,R)^{-1}$, then $c_1=O(1)$ and $\la c_1
>0$.
\end{lemma}

This lemma can be proved using standard perturbation argument
and similar conclusions hold for excited states as well.
For the purpose of this paper, we prefer to
use the value $m=(\phi_1, \,
Q_1)$ as the parameter and refer to the family of excited states
as $Q_1(m)$. It is straightforward to compute the leading
corrections of $Q_1(m)$ via standard perturbation argument
used in proving Lemma \ref{th:2-1}.  Thus we can  write $Q_1$ as
\begin{equation}
Q_1(m)= m \phi_1 + \bke{m^3 q_3 + q^{(5)}(m)}: =   m \phi_1 + q(m)
\; , \quad q^{(5)}(m)=O(m^5), \quad q(m) \perp \phi_1
~, \label{eq:2-2}
\end{equation}
where $q_3 = - \la (H_0 - e_1)^{-1} \pi \phi_1^3$
and  $\pi$ is  the projection
\begin{equation} \label{pi.def}
\pi h = h - (\phi_1, h) \phi_1 .
\end{equation}
Similarly, we can also expand  $E_1(m)$ in $m$ as
\begin{equation} \label{E.dec}
  E_1(m)= e_1 + E_{1,2} m^2 + E_{1,4}m^4 + E_1^{(6)}(m),
  \qquad E_1^{(6)}(m)= O(m^6) ~.
\end{equation}
Moreover, we can differentiate the  relation of $Q_1(m)$
w.r.t. $m$ to have
\begin{equation}
 Q_1'(m)= \frac d{d m}Q_1 =\phi_1 + q'(m) ~, \qquad
q'(m)= \frac
d{dm}q(m) = O(m^2),   \quad q'(m) \perp
 \phi_1,
\end{equation}

\subsection{Equations}

Thus in the first and second time regimes, we write
\begin{equation}\label{eq:2-4}
\psi(t) = x(t)\phi_0 + Q_1(m(t))e^{i \Theta(t)} + \xi(t) ~,
\end{equation}
where $\xi \in \Hc(H_0)$, see \eqref{psidec1}. If we
write $\Theta(t)=\theta(t) - \int_0^t E_1(m(s)) \, ds$, we
can write $y(t)$ as
\begin{equation} \label{y.def}
y= m e^{i \Theta} = m \exp \bket{i\theta(t) - i\int_0^t E_1(m(s)) \,
ds} ~.
\end{equation}

Denote the part orthogonal to $\phi_1$ by $h ={x}\phi_0 +{\xi}$.
From the Schr\"odinger
equation, we have $h$ satisfies the equation
\begin{align}
i \pd_t h &= H_0 h + G + \Lambda, \nonumber
\\
G &= \la |\psi|^2 \psi - \la Q_1^3 e^{i \Theta} \nonumber \\
&=\la Q_1^2(e^{i2\Theta} \bar h +2 h ) + \la Q_1(e^{i\Theta}
2h\bar h +e^{-i\Theta} h^2 ) + \la |h|^2 h
\\
\Lambda &=  \bke{\dot \theta Q_1 - i \dot m Q_1' } e^{i \Theta} ~.
\end{align}
Since  $m(t)$ and $\theta(t)$ are chosen so that \eqref{eq:2-4}
holds, we have  $0 = (\phi_1, i\pd_t h(t))
= \bke{ \phi_1, \, G + (\dot \theta Q_1 - i \dot m Q_1')e^{i
\Theta} } $. Hence $m(t)$ and $\theta(t)$ satisfy
\begin{equation} \label{mt.eq}
\dot m = \bke{ \phi_1, \Im G e^{-i \Theta} } \, , \qquad \dot
\theta = -\frac 1m \, \bke{ \phi_1, \Re G e^{-i \Theta}} ~.
\end{equation}

We also have the equation for $y$:
\[
i \dot y  = i \dot m e^{i \Theta} - ( \dot \theta- E_1(m) ) m e^{i
\Theta} = E_1(m)y + e^{i \Theta} (i \dot m - m \dot \theta ) = E_1(m)y +
(\phi_1 , G)~.
\]
Here we have used \eqref{mt.eq}.  Denote
$\Lambda_\pi= \pi \Lambda$. We can decompose the equation
for $h$ into equations for $x$ and $\xi$. Summarizing, we have
the original Schr\"odinger equation is equivalent to
\begin{equation}
\label{xyxi.eq} \left \{
\begin{aligned}
i\dot x &= e_0 \, x + \bke{ \phi_0, \, G + \Lambda_\pi }
\\
i \dot y & = E_1(m)y + (\phi_1 , G)
\\
i \pd_t \xi &= H_0 \, \xi + \Pc \bke{ G + \Lambda_\pi }
\end{aligned} \right .
\end{equation}

Clearly, $x$ has an oscillation factor $e^{-ie_0t}$, and $y$ has
a factor $e^{-ie_1t}$. Hence we define
\begin{equation} \label{eq:uv}
  x= e^{-i e_0 t} u, \qquad y=e^{-i e_1 t} v ~.
\end{equation}
Together with the integral form of the equation for $\xi$, we have
\begin{align}
\dot u &= -i e^{ie_0t}  \bke{ \phi_0, \, G + \Lambda_\pi }
,\label{u.eq}
\\
\dot v &= -i e^{ie_1 t} \bkt{(E_1(m)-e_1)y + (\phi_1 , G)},
\label{v.eq}
\\
\xi(t) &= e^{-i H_0  t}\xi_0  +\int_0^t e^{-i H_0 (t-s)} \, \PcH
G_\xi(s)\, d s ~,\qquad G_\xi = i^{-1}(  G + \Lambda_\pi)
~.\label{xi.eq}
\end{align}
This is the system we shall study.

\subsection{Basic estimates and decompositions}

It is useful to decompose various terms according to orders in $n$
so that we can identify their contributions. We now proceed
to so this for  $G$, $\Lambda_\pi$, $E(|y|)$ and $\xi(t)$.
We expect that $x,y=O(n)$ and $\xi=O(n^3)$ locally.

{\bf 1. $G$}\quad 
Recall that $G$ is given by
\[
G=  \la Q_1^2 (e^{i2\Theta} \bar h +2 h ) + \la
Q_1 (e^{i\Theta} 2h\bar h +e^{-i\Theta} h^2 ) + \la |h|^2 h
\]
with $h=x\phi_0 + \xi$ and $Q_1=Q_1(|y|)$. From
the the decomposition  \eqref{eq:2-2} of  $Q_1=|y|\phi_1 + |y|^3 q_3
+ q^{(5)}(|y|)$, we  decompose $G$ as
\begin{align*}
G&=\la (y^2 \phi_1^2 + 2 y^3 \bar y \phi_1 q_3) \, \bar h + \la
(|y|^2 \phi_1^2 + 2 |y|^4 \phi_1 q_3) \,  2h
\\
&\quad + \la (y \phi_1 + y^2 \bar y q_3 )\, 2 |h|^2 + \la (\bar y
\phi_1 + y \bar y^2 q_3 ) \, h^2 + \la |h|^2 h + (*)
\end{align*}
where $(*)=\la \bkt{2|y| \phi_1 q^{(5)} + (|y|^3 q_3 + q^{(5)})^2}
\, (e^{i2\Theta} \bar h +2 h ) + \la q^{(5)}\, (e^{i\Theta} 2h\bar
h +e^{-i\Theta} h^2 )$ with $q^{(5)}=q^{(5)}(|y|)$. We then
substitute $h=x\phi_0 + \xi$ to obtain
\begin{equation} \label{G.dec}
G= G_3 + G_5 + G_7,
\end{equation}
where
\begin{align}
\label{G3.def} G_3 &=  \la (y^2 \bar x + 2|y|^2 x) \phi_0 \phi_1^2
  + \la (2|x|^2 y + x^2 \bar y) \phi_0^2 \phi_1 +\la |x|^2 x \phi_0^3
\\
\label{G5.def} G_5 &=\la (2 y^3 \bar y \bar x   + 4 |y|^4 x ) \,
\phi_0\phi_1 q_3  + \la (2|x|^2 y^2 \bar y + x^2 y \bar y^2 ) \,
\phi_0^2 q_3
\\  \nonumber
& \quad +  \la (x \phi_0 + y \phi_1)^2 \wbar \xi + 2\la |(x \phi_0
+ y \phi_1)|^2 \xi
\end{align}
and
\begin{align} \label{G7.def}
 G_7 &=  \la \bkt{2|y| \phi_1 q^{(5)}(|y|) + (|y|^3 q_3 + q^{(5)}(|y|))^2}
\, (e^{i2\Theta} \bar h +2 h )
\\
\nonumber &\quad + \la  q^{(5)}(|y|)\, (e^{i\Theta} 2h\bar h
+e^{-i\Theta} h^2 )
\\  \nonumber
& \quad +  \la \bke{  2 y^3 \bar y \phi_1 q_3 \, \bar \xi +   4 |y|^4
\phi_1 q_3 \,  \xi }
 + \la (y \phi_1 2 |\xi|^2 + \bar y \phi_1 \xi^2)
\\ \nonumber
&\quad  +\la  y^2 \bar y q_3 \, 2 ( x\phi_0 \bar \xi + \bar x
\phi_0 \xi + |\xi|^2) + \la y \bar y^2 q_3  \, (2x \phi_0 \xi +
\xi^2)
\\
\nonumber &\quad + \la \phi_0(\bar x \xi^2 + 2 x |\xi|^2  ) + \la
|\xi|^2 \xi ~.
\end{align}

Note that $G_3=O(n^3)$, $G_5=O(n^5)$ and $G_7=O(n^7)$.
If we use the convention that
\[
f \leC g_1 + g_2 + \cdots
\]
for  $\norm{f} \le C \norm{g_1}+\norm{g_2}+ \cdots$ for some
suitable norms, we have
\begin{align}
G &\leC n^2 x + n^2 \xi + \xi^3 \label{G.est0}\\
G_5 &\leC n^4 x + n^2 \xi \label{G5.est0}\\
G_7 &\leC n^6 x + n^4 \xi + n \xi^2 + \xi^3 \label{G7.est0}
\end{align}

It is crucial to observe that {\it no term in $G_3$ is of order
$y^3$}. This is due to  our setup emphasizing the role of
nonlinear excited states.
The price we pay is the introduction of terms involving
$q_3$ and $q^{(5)}$.

We now  identify the main  oscillation factors of various
terms. For example, $y^2 \bar x= e^{i(-2e_1+e_0) t} v^2 \bar u$,
and its factor is $-2e_1+e_0$. For terms in $G_3$ we have
\begin{equation} \label{eq:2-18}
\begin{array}{c} y^2 \bar x \\ -2e_1 + e_0 \end{array} \quad
\begin{array}{c} |y|^2 x \\  -e_0  \end{array} \quad
\begin{array}{c} |x|^2 y\\ -e_1  \end{array} \quad
\begin{array}{c} x^2 \bar y \\ -2 e_0 + e_1  \end{array} \quad
\begin{array}{c} |x|^2 x \\-e_0  \end{array} \quad
\end{equation}
From the spectral assumption  $|e_0|
> 2|e_1|$, $-2e_1 + e_0$ is the only  negative phase
factor. Hence it is the only term of order $(n^3)$ that have
resonance effect when we compute the main part of $\xi$. Also,
since $|x|^2 y$ has the same phase as $y$, it will be resonant in
the $y$-equation. Similarly, $|y|^2x$ and $|x|^2x$ have same phase
as $x$ and will be resonant in $x$-equation.

{\bf 2. $\Lambda_\pi$ and $E(m)$}\quad
Recall $\Lambda_\pi= \pi (\dot \theta Q_1 - i \dot m Q'_1) e^{i
\Theta}$. Since $\dot \theta = O(n^{-1}\norm{G}\loc)$ and $\dot
m=O(\norm{G}\loc)$,
\begin{equation} \label{La.est}
\norm{\Lambda_\pi(s)} = O(\dot \theta)\,O(\pi Q_1) + O(\dot
m)\,O(\pi Q_1') \le C n^2 \norm{G}\loc ~,
\end{equation}

To find out the main part of $\Lambda_\pi$, we substitute equation
\eqref{mt.eq} for $\dot m$ and $\dot \theta$ to obtain ( $m=|y|$),
\begin{align*}
\Lambda_\pi &=  \pi (\dot \theta Q_1 - i \dot m Q'_1 )\, e^{i
\Theta} \\
&= -\bket{(\phi_1,\, G/2)m^{-1} \pi Q_1 + (\phi_1,\, \bar G/2)
m^{-1}  \pi Q_1 e^{2i\Theta}}\\
&\quad  -i\bket{(\phi_1,\, G/2i) \pi Q'_1 + (\phi_1,\, \bar G/2i)
\pi Q'_1 e^{2i\Theta}}
\end{align*}
Since $G=G_3 + (G_5+G_7)$ and $\pi Q_1(m)=m^3 q_3 + q^{(5)}(m)$ by
\eqref{eq:2-2}, we have $\pi Q'_1(m)=3 m^2 q_3 + O(m^4)$, and the
main part of $\Lambda_\pi$ is (also recall $y=m e^{i \Theta}$)
\begin{align}
\Lambda_{\pi,5} &= -\tfrac 12 \bket{(\phi_1,\, G_3)\,|y|^2 q_3
+ (\phi_1,\, \bar G_3) \,y^2 q_3}\nonumber\\
&\quad  -\tfrac 12 \bket{(\phi_1,\, G_3) \,3|y|^2 q_3 + (\phi_1,\,
\bar G_3) \,3y^2 q_3}\nonumber \\
&=- 2 q_3 \, (\phi_1,\, G_3 |y|^2 + \bar G_3 y^2) \label{La5.def}
\end{align}
Let $\Lambda_{\pi,7}=\Lambda_{\pi}-\Lambda_{\pi,5}$. We have
\begin{align}
\Lambda_{\pi} &= \Lambda_{\pi,5}+\Lambda_{\pi,7} \label{La.dec}\\
\Lambda_{\pi,5} &\leC \norm{G_3}\loc |y|^2 \leC n^4 x \label{La5.est}\\
\Lambda_{\pi,7} &\leC \norm{G_5+G_7}\loc |y|^2 + \norm{G}\loc
|y|^4 ~. \label{La7.est}
\end{align}

{\bf 3. $\xi$}\quad Recall the equation for $\xi$ in
\eqref{xi.eq},
\[\xi(t) = e^{-i H_0  t}\xi_0  +\int_0^t e^{-i H_0
(t-s)} \, \PcH G_\xi(s)\, d s ~,\qquad G_\xi = i^{-1}(  G +
\Lambda_\pi) ~.
\]
Since $\norm{\Lambda_\pi} \le C n^2 \norm{G}\loc $, the main terms
in $G_\xi=i^{-1}(G+\Lambda_\pi)$ is $i^{-1}G_3$.
We now compute the first term
$\la y^2 \bar x \phi_0 \phi_1 ^2$ in $G_3$ using integration by
parts:
\begin{align*}
& - i \la \int_0^t e^{-i H_0  (t-s)} \, \Pc y^2 \bar x \phi_0
\phi_1 ^2 \, d s
\\
&= - i \la e^{-i H_0  t} \int_0^t  e^{i (H_0 -0i) s} \, e^{i
(e_0-2e_1) s} \, v^2 \bar u \Pc \phi_0 \phi_1 ^2 \, d s
\\
&=- i \la e^{-i H_0  t} \Bigg \{  \bkt{ \frac 1{ i (H_0 -0i +e_0-
2e_1)} e^{i H_0  s} \, e^{ i (e_0-2e_1) s} \, v^2 \bar u \Pc
\phi_0\phi_1 ^2 }_0^t
\\
& \qquad \qquad \qquad - \int_0^t \frac 1{ i (H_0 -0i +e_0- 2e_1)}
e^{i H_0  s} \, e^{i (e_0-2e_1) s} \,
 \frac d{d s} \bke{ v^2 \bar u} \Pc \phi_0 \phi_1^2 \, d s \Bigg \}
\\
&= y^2 \bar x \Phi_1 - e^{-i H_0  t} y^2 \bar x(0) \Phi_1 -
\int_0^t e^{-i H_0 (t-s) } \, e^{i (e_0-2e_1) s} \frac d{d s}
\bke{ v^2 \bar u} \Phi_1  \, d s ~,
\end{align*}
where
\begin{equation}\label{Phi1.def}
\Phi_1 = \frac {-\la} { H_0 -0i +e_0- 2e_1} \, \Pc \phi_0 \phi_1
^2 ~.
\end{equation}
This term,  with the phase factor  $e_0 - 2 e_1$,  is the only
one in $G_3$ having  a negative phase factor (see
\eqref{eq:2-4}). Since $-(e_0 - 2 e_1)$ is in  the continuous
spectrum of $H_0 $, $H_0 +e_0- 2e_1$ is not invertible, and needs
a regularization  $-0i$. We choose
$-0i$, not $+0i$, so that the term $e^{-i H_0  t} y^2 \bar x(0)
\Phi_1$ decays as $t \to \infty$. See Lemma \ref{th:2-2}.

We can integrate all terms in $G_3$ and obtain  the main terms of
$\xi(t)$ as
\begin{equation}
\xitwo(t)= y^2 \bar x \Phi_1 + |y|^2 x \Phi_2 + |x|^2 y \Phi_3 +
x^2 \bar y \Phi_4  + |x|^2 x \Phi_5
\end{equation}
where
\begin{alignat}{2} \label{Phij.def}
\Phi_2 &=  \frac {-2\la} { H_0  -e_0 } \, \Pc \phi_0 \phi_1 ^2 ~,
\qquad &\Phi_3& =  \frac {-2\la} { H_0  -  e_1} \, \Pc \phi_0^2
\phi_1 ~,
\\
\Phi_4 &=  \frac {-\la} { H_0   -2 e_0 + e_1} \, \Pc \phi_0 ^2
\phi_1 ~, \qquad &\Phi_5 &=  \frac {-\la} { H_0   - e_0} \, \Pc
\phi_0 ^3 ~. \nonumber
\end{alignat}
The rest of $\xi(t)$ is
\begin{align*}
\xithree(t)&= e^{-i H_0  t} \xi_0 -e^{-i H_0  t}\xitwo(0)
-\int_0^t e^{-i H_0 (t-s)} \, \Pc \, G_4 \, d s
\\
&\quad + \int_0^t e^{-i H_0 (t-s)} \, \Pc
\bke{G_\xi-i^{-1}G_3-i^{-1} \la|\xi|^2\xi} \, d s \\
& \quad + \int_0^t e^{-i H_0 (t-s)} \, \Pc
\bke{i^{-1}\la|\xi|^2\xi}\, d s
\\
&\equiv \xithree_1(t)+ \xithree_2(t) +\xithree_3(t)
+\xithree_4(t)+ \xithree_5(t)
\end{align*}
where
$\xithree_4(t)$ and $\xithree_5(t)$ are higher order terms in
$G_\xi$ which we did not integrate and
the integrand $G_4$ in
$\xithree_3(t)$ consists of the remainders from the integration by
parts:
\begin{align}
G_4 &= e^{i(e_0 - 2 e_1)  s} \frac d{d s} \bke{ v^2 \bar u} \Phi_1
+ e^{i(-e_0 )  s} \frac d{d s} \bke{ |v|^2 u} \Phi_2
\label{G4.def}
\\
& \quad + e^{i ( -  e_1)  s} \frac d{d s} \bke{ |u|^2  v} \Phi_3 +
e^{i (-2 e_0 + e_1)  s} \frac d{d s} \bke{ u^2 \bar v} \Phi_4 +
e^{i (-e_0 )  s} \frac d{d s} \bke{ u^2 \bar u} \Phi_5 ~,\nonumber
\end{align}
Here we single out $\xithree_5(t)$
since $|\xi|^2\xi$ is a non-local term. Thus
we have following   decomposition for $\xi$:
\begin{equation}\label{xi.dec}
\xi(t)= \xitwo(t)+\xithree(t) = \xitwo +\bke{\xithree_1 + \cdots +
\xithree_5}~,
\end{equation}

\subsection{Linear estimates}

We now summarize known results concerning  the linear analysis.
The decay estimate was contained in  \cite{JSS} and \cite{Y};
the estimate \eqref{eq:22-1B} was taken from \cite{SW2} and  \cite{TY};

\begin{lemma}[decay estimates for $e^{-itH_0}$] \label{th:2-2}
For $q \in [2, \infty]$ and $q'=q/(q-1)$,
\begin{equation}  \label{eq:22-1A}
\norm{  e^{-itH_0} \, \Pc^{H_0} \phi }_{L^q} \le C \,|t|^ {-3
\bke{\frac 12 - \frac 1q}} \norm{\phi}_{L^{q'}} ~.
\end{equation}
For smooth local functions $\phi$ and sufficiently large $r_0$, we
have
\begin{equation} \label{eq:22-1B}
\lim_{\sigma\to 0+} \norm{ \bkA{x}^{-r_0} \, e^{-it H_0} \,
\frac 1{(H_0  + e_0 - 2 e_1- \sigma i)^k}
\Pc^{H_0} \bkA{x}^{-r_0} \phi }_{L^2} \le C \bkA{t}^{-9/8}
\end{equation}
where $k=1,2$.
\end{lemma}

Intuitively, we can write
\[
e^{-i H_0  t} \Phi_1 = \lim_{ \e \to 0+ } e^{-i H_0  t} \int_0^\infty
e^{-i (H_0 - \e i + e_0 - 2 e_1)s} \Pc \phi_0 \phi_1 ^2 \, ds ~.
\]


\section{The initial layer and the transition regimes: The estimates}

In this section we wish to show the following picture
for the  solution $\psi(t)$:
In the initial layer regime, the dispersive part gradually disperses
away, while the sizes of the bound states do not change much.
In the transition regime, the {\it original} dispersive part becomes
negligible, while the $\phi_0$-components of $\psi(t)$ increases
and the $\phi_1$-component decreases.

Recall the orthogonal decomposition
$\psi(t)=\lbar{x}\phi_0 + \lbar{y}\phi_1 + \lbar{\xi}$
\eqref{psidec0}.
We have $ |\lbar{x}(t)|^2 + |\lbar{y}(t)|^2 +
\norm{\lbar{\xi}(t)}_{L^2}^2 =\norm{\psi(t)}_{L^2}^2 \le \n^2  $.
If we  decompose $\psi$ via  \eqref{psidec1}, i.e.,   
\begin{equation}
\psi(t)=x\phi_0 + Q_1(y) + \xi ~, \qquad
\xi= \xitwo + \xithree ~ ,
\end{equation}
we have
%
%
$y=\lbar{y}$,
$x=\lbar{x}+O(y^3)$ and $\xi=\lbar{\xi}+O(y^3)$. Thus
\begin{equation}  \label{eq:3-2}
|x(t)|, \; |y(t)|, \; \norm{\xi(t)}_{L^2} \le \tfrac 54\n ~, \qquad
 \norm{\xi_0}_{Y} \le 4\n ~.
\end{equation}

For $p=1,2,4$, define the space
\begin{equation}\label{eq:3-3}
{L^p \loc } \equiv  \bket{f: \; \bkA{x}^{-r_0}f(x) \in L^p(\R^3)} ~,
\end{equation}
where $r_0>3$ is the exponent appeared in the linear estimate
\eqref{eq:22-1B} in Lemma \ref{th:2-2}.
Since $r_0>3$,  we have
$\norm{f}_{L^1 \loc} \le C\norm{f}_{L^p}$ for any $p$,

The following proposition is the main result for the dynamics in
the  initial layer and the transition regimes.

\begin{proposition}  \label{th:3-1}
Suppose that $V$ satisfies the  assumptions  given in \S 1.
Let $\psi(t,x)$ be
a solution of \eqref{Sch} with the initial data $\psi_0$ satisfying
\eqref{eq:1-10}. Let $\e_3>0$ be a sufficiently
small constant  to be fixed later. Let $t_0=\e_3 n^{-4}$.
Then there exists $t_1$ and $t_2$ such that
for some constant $C \le 10000$
we have
\begin{equation}
\label{eq:3-4}
t_0\le  t_1\le \frac {1.01}{\;\gamma_0
\n^4\;}\; \log \bke{\frac \n {|x_0|}} ~, \qquad
 t_1 + C \; (\n^4 \e^2)^{-1}\le  t_2\le t_1 +
10100\; (\gamma_0 \n^4 \e^2)^{-1} ~,
\end{equation}
and the following estimates hold:

(i) For $0\le t\le 2\,t_2$,
\begin{equation} \label{eq:3-5}
|x(t)|\ge \tfrac 34 \,\sup_{0\le s\le t} \, |x(s)| ~,
\end{equation}
\begin{align}
\norm{\xi(t)}_{L^4} &\le C_2\, \n^2 t^{1/4}\,|x(t)|
+ C_2 \norm{\xi_0}_Y \bkA{t}^{-3/4} ~,
\nonumber
\\
\norm{\xi(t)}_{L^4\loc} &\le C_2 n^2 \,|x(t)| \quad
+ C_2 \norm{\xi_0}_Y \bkA{t}^{-9/8} ~,
\label{xi.est}
\\
\norm{\xithree(t)}_{L^2 \loc} &\le  \, C_2\, \n^{15/4} |x(t)|
+ C_2 \norm{\xi_0}_Y \bkA{t}^{-9/8} ~.
\nonumber
\end{align}
where the constant $C_2$ will be specified in \eqref{eq:C2} of next
subsection.

(ii) (Initial layer) for $0\le t \le t_0$,
\begin{equation} \label{eq:3-7}
\begin{array}{l}
\tfrac 12 |x_0| \le |x(t)| \le \tfrac 32\, |x_0|,
\\
0.99\,|y_0| \le |y(t)|\le 1.01\,|y_0|
\end{array}
\end{equation}
(iii) For $n_1=\bke{|x_0|^2 + \tfrac 12 \,
|y_0|^2}^{1/2}$ defined in \eqref{n1.def},
\begin{equation} \label{eq:3-8}
  |x(t_1)| \ge 0.01 \n ~, \quad |x(t_2)| \ge 0.99 \n_1 ~, \quad
\tfrac 12 \e n \le |y(t_2)| \le 2\e\n ~.
\end{equation}
\end{proposition}

\bigskip

Notice that \eqref{eq:3-4} implies
\begin{equation} \label{eq:3-4B}
t_2 \le C_3 n^{-4} 
\end{equation}
for some constant $C_3$.

We will prove these estimates using \eqref{eq:1-10},
\eqref{eq:3-2} and a continuity argument. Hence we can assume
the following weaker estimates:  For $0\le t\le
2t_2$:
\begin{equation}
\label{A:pf}
\begin{aligned}
|x(t)|&\ge \tfrac 12 \,\sup_{0\le s\le t} \, |x(s)| ~,
\\
|x(t)|&\le 2 |x_0| \qquad \text{for } t< t_0 ~,
\\
\norm{\xi(t)}_{L^4} &\le 2 C_2\, \n^2 t^{1/4}\,|x(t)|
+ 2 C_2 \norm{\xi_0}_Y \bkA{t}^{-3/4} ~,
\nonumber
\\
\norm{\xi(t)}_{L^4\loc} &\le 2 C_2 n^2 \,|x(t)| \quad
+ 2 C_2 \norm{\xi_0}_Y \bkA{t}^{-9/8} ~,
\\
\norm{\xithree(t)}_{L^2 \loc} &\le  \, 2 C_2\, \n^{15/4} |x(t)|
+ 2 C_2 \norm{\xi_0}_Y \bkA{t}^{-9/8} ~.
\nonumber
\end{aligned}
\end{equation}
By continuity, if we prove Proposition \ref{th:3-1} assuming
these weaker estimates, we have proved the proposition itself.
We shall see also estimates \eqref{A:pf} will  be used only in estimating
higher order terms.

Recall from \eqref{La.est} that the local term $\Lambda_\pi$
satisfies $\norm{\Lambda_\pi}_r\le C n^2 \norm{G}_{L^1 \loc}$ for any
$r$. Thus we have
\begin{equation} \label{uvdot.est}
|\dot u(t)| \leC \norm{G}_{L^1 \loc}, \qquad |\dot v(t)| \leC
\norm{G}_{L^1 \loc} + |y|^3 .
\end{equation}

The following lemma provides estimates for $G$ assuming
the  estimate \eqref{A:pf}.

\begin{lemma} \label{th:3-2}
Let $G$ be given by \eqref{G.dec}--\eqref{G7.def}.
Suppose   $n$ is sufficiently small
and the  estimate \eqref{A:pf}
holds  for  $t\le C_3 n^{-4}$. 
Then we  have the following estimates for $G$:
\begin{equation} \label{G.est}
\norm{G(t)}_{L^{4/3}\cap L^{8/7}}
\le C_4 \n^{2}|x(t)| + C(C_2) n^2 \norm{\xi_0}_Y \bkA{t}^{-9/8}~.
\end{equation}
\begin{equation} \label{G57.est}
\norm{(G-G_3)(t)}_{L^{4/3}\cap L^{8/7}}
\le C_4 \n^{15/4}|x(t)|+ C(C_2) n^2 \norm{\xi_0}_Y \bkA{t}^{-9/8}~.
\end{equation}
\begin{equation} \label{G57.estB}
\norm{(G-G_3)(t)}_{L^1 \loc} \le C_4 \n^{4}|x(t)|+ C(C_2)
n^2 \norm{\xi_0}_Y \bkA{t}^{-9/8}~.
\end{equation}
where $C_4$ is a constant independent of $C_2$ and  $C(C_2)$
denotes constants depending on $C_2$.

Moreover, \eqref{G.est} and \eqref{G57.est} remain true if we
replace $G$ by $G_\xi$, and $(G-G_3)$ by  $(G_\xi-i^{-1}G_3)$.
Furthermore, we have
\begin{equation} \label{Gxi.L1est}
\norm{G_\xi(t)}_{L^{1}}\le C_5 n^3 ~.
\end{equation}

\end{lemma}

By the assumption \eqref{eq:1-10}, when $t>t_0$ the last
term $C n^2 \norm{\xi_0}_Y \bkA{t}^{-9/8} $ is smaller and can
be removed. The proof of this lemma is a straightforward application of
the Holder and Schwarz inequalities.

We will use \eqref{G.est} and \eqref{G57.est} for $\xi$,
and \eqref{G57.estB} for $x$ and $y$.

\myproof
Recall \eqref{G.dec} that $G=G_3+G_5+G_7$.
We first consider  only the non-local term $\la |\xi|^2\xi$ in $G$,
which belongs to $G_7$. Since $n\le \e^2$ and $t_2 \le C_3 \e^{-2}
\n^{-4}$, by \eqref{A:pf} we have $\norm{\xi(s)}_{L^{4}}\le C n
^{3/4}|x(s)|+  C \norm{\xi_0}_Y \bkA{s}^{-3/4}$. Also using
\eqref{eq:3-2} and the H\:older inequality we have
\begin{equation} \label{eq:3-21}
 \norm{|\xi|^2\xi(s)}_{L^{4/3}}\le C \norm{\xi(s)}_{L^{4}}^3
\le C \bke{n ^{3/4}|x(s)|}^3  + C \norm{\xi_0}_Y^3 \bkA{s}^{-9/4}
~.
\end{equation}
\begin{align}
\norm{|\xi|^2\xi(s)}_{L^{8/7}}&\le C \norm{\xi(s)}_{L^{2}}^{1/2}
\, \norm{\xi(s)}_{L^{4}}^{5/2} \nonumber
\\
&\le C \n^{1/2}\,\bket{\bke{\n^{3/4} |x(s)|}^{5/2} +
\norm{\xi_0}_Y^{5/2} \bkA{s}^{-15/8} }\label{eq:3-21A}
\\
&\le C \n^{4 -1/8}|x(s)|+ C \n^{2}\,\norm{\xi_0}_Y
\bkA{s}^{-15/8}~.\nonumber
\end{align}
Hence this non-local term satisfies \eqref{G.est}--\eqref{G57.est}.
Moreover, to prove \eqref{G57.estB}, we can bound
$\norm{|\xi|^2\xi(s)}_{L^1 \loc}$ by
$\norm{\xi(s)}_{L^{4}}^3$.

For the local term $G -\la|\xi|^2\xi= G_3 + G_5 +
(G_7-\la|\xi|^2\xi)$, all $L^p$-norms
are equivalent. We can read from the explicit expressions
of $G$  the
following estimates:
\begin{align*}
G_3 &\leC n^2 x
\\
G_5 & \leC n^4 x + \n^2\,\xi
\\
G_7-\la|\xi|^2\xi& \leC \n^6\,\xm+ \n^4\,\xi + \n\,\xi^2
\end{align*}
To estimate $\xi$ in $G_\xi -\la|\xi|^2\xi$, we can use
$\norm{\xi}_{L^{4}\loc}$. For example,
\[
\norm{\bar y\phi_1\xi^2}_{L^{4/3}}\le
C|y|\norm{\phi_1\bkA{x}^{2r_0}}_{L^{4}}\,\norm{\xi}_{L^{4}\loc}^2
\le C n \bke{(n^2 \xm)^2 +\norm{\xi_0}_Y^2
\bkA{s}^{-9/4}}~.
\]
Together with  the explicit expressions
of $G$ and $G_3$,  similar arguments  show \eqref{G.est}--\eqref{G57.estB}.

Since $G_\xi= i^{-1}(G+\Lambda_\pi)$ and
$\norm{\Lambda_\pi}\le Cn^2 \norm{G}\loc$ by \eqref{La.est},
\eqref{G.est} and \eqref{G57.estB} hold if we replace $G$ and
$G-G_3$ by $G_\xi$ and
$G_\xi- i^{-1}G_3$.
obtain \eqref{Gxi.L1est}, we only need to check the non-local term
$\la |\xi|^2\xi$. Since $\norm{\xi(s)}_{L^{4}}\le (2C_2C_3^{1/4}+8C_2)n$, we have
\begin{equation} \label{eq:3-20A}
\norm{|\xi|^2\xi}_{L^{1}}\le C_{4,1} \norm{\xi(s)}_{L^{2}} \,
\norm{\xi(s)}_{L^{4}}^{2} \le \frac 1{10}C_5 \n\, n^2 ~,
\end{equation}
provided we choose  $C_5 \ge 10(2C_2C_3^{1/4}+8C_2)C_{4,1}$.
Thus the lemma is proved. \myendproof

\subsection{Estimates of the dispersive part}

We now prove the estimates for $\xi$ in Proposition
\ref{th:3-1} by using \eqref{eq:3-2}, \eqref{A:pf},
Lemma \ref{th:2-2} and Lemma \ref{th:3-2}.


\bigskip

{\bf Step 1.} $L^4$ and $L^4\loc$ norms, $0\le t \le 2t_2$

Recall  the equation \eqref{xi.eq} for $\xi$:
\begin{equation} \label{E3-21}
\xi(t) = e^{-i H_0  t}\xi_0  +\int_0^t e^{-i H_0 (t-s)} \, \PcH
G_\xi(s)\, d s ~,\qquad G_\xi = i^{-1}(  G + \Lambda_\pi) ~.
\end{equation}
By \eqref{A:pf},
Lemma \ref{th:2-2} and Lemma \ref{th:3-2}, we have
\begin{align*}
\norm{\xi(t)}_{L^4} &\le \norm{ e^{-i H_0  t}\xi_0 }_{L^4} +
\int_0 ^{t} C \length{t-s}^{-3/4} \norm{G_\xi(s)}_{4/3} \, d s
 \\
&\le C \norm{\xi_0}_Y \bkA{t}^{-3/4} +  \int_0 ^{t} C
\length{t-s}^{-3/4} \bke{C_4\n^2 |x|(s) + C(C_2) n^2 \norm{\xi_0}_Y
\bkA{s}^{-9/8} } \, d s
 \\
&\le C_{2,1} \norm{\xi_0}_Y \bkA{t}^{-3/4} + C_{2,1} \n^2
\,t^{1/4}\, |x(t)| \, + \, C(C_2) n^2 \norm{\xi_0}_Y
\bkA{t}^{-4/3},
\end{align*}
where $C_{2,1}$ is some  explicit  constant.

We now estimate $\norm{\xi(t)}_{L^4 \loc}$. If $t\le 1$, we can
bound $L^4 \loc$-norm by $L^4$-norm. Hence we may assume $t>1$.
For ,
We divide the time integral in \eqref{E3-21} into
$s\in [0,t-1]$ and  $s\in [t-1,t]$. In the first interval,
we estimate $L^4 \loc$ by $L^8$ norm; in the second we
estimate $L^4 \loc$ simply  by $L^4$. Using similar arguments  in
the previous estimate of the $L_4$ norm,  we  have
\begin{align*}
\norm{\xi(t)}_{L^4 \loc} &\le C \norm{\xi_0}_Y \bkA{t}^{-9/8} +
\int_0^{t-1} \frac C{|t-s|^{9/8}} \norm{G_\xi}_{L^{8/7}} \, d s +
\int_{t-1}^t \frac C{|t-s|^{3/4}} \norm{G_\xi}_{L^{4/3}} \, d s
\\
&\le C \norm{\xi_0}_Y \bkA{t}^{-9/8} + \int_0^t \frac C
{\bkA{t-s}^{9/8}}
\bke{C_4\n^2 |x|(s) + C(C_2) n^2 \norm{\xi_0}_Y
\bkA{s}^{-9/8} } \, d s
\\
&\qquad + \sup_{t-1 \le s \le t} \norm{G_\xi(s)}_{L^{4/3}}
\\
&\le C_{2,2} \norm{\xi_0}_Y \bkA{t}^{-9/8} + C_{2,2} n^2 |x(t)|
\, + \, C(C_2) n^2 \norm{\xi_0}_Y
\bkA{t}^{-9/8}
\end{align*}
where $C_{2,2}$ is some  explicit  constant.

\bigskip

{\bf Step 2.} $L^2 \loc $-norm, $0\le t \le 2t_2$

Recall the decomposition \eqref{xi.dec}: $\xi=\xitwo+\xithree=
\xitwo +(\xithree_1 + \cdots + \xithree_5)$. We will estimate the
$L^2 \loc $-norm of each term.

{\bf 0}. $\xitwo$. Since $\Phi_1 \in L^2 \loc$, and $\Phi_j\in
L^2$, $(j>1)$, we have
\[
\norm{\xitwo(t)}_{L^2 \loc} \le C_{2,3} C\n^2 \, |x(t)| ~,
\]
for some explicit constant $C_{2,3}$.

{\bf 1}. $\xithree_1$. We have
\[
\norm{\xi^{(3)}_1(t)}_{L^2 \loc}\le C_{2,4} \norm{\xi_0}_Y
\bkA{t}^{-9/8}
 ~,
\]
for some explicit constant $C_{2,4}$ by the $L^{p',p}$ estimate of
$e^{-itH_0}$ in Lemma \ref{th:2-2}.

{\bf 2}. $\xithree_2$. By the linear estimate \eqref{eq:22-1B} in Lemma
\ref{th:2-2} we have, for some constant $C_{2,5}$,
\[
\norm{\xithree_2(t)}_{L^2 \loc}\le C_{2,5} \n^2 |x_0| \bkA{t}^{-9/8}
~.
\]

{\bf 3}. $\xithree_3$. To estimate $\xithree_3(t)=-\int_0^t
e^{-iH_0(t-s)} \Pc G_4 \, ds$ with $G_4$ defined in \eqref{G4.def},
we need estimates \eqref{uvdot.est} for $\dot u$ and $\dot v$
and the linear estimate \eqref{eq:22-1B} in Lemma \ref{th:3-2}. Hence
\begin{align*}
\norm{\xithree_3(t)}_{L^2 \loc}& \le
\int_0^t \norm{e^{-iH_0(t-s)} \Pc G_4}_{L^2 \loc} \, ds
\\
&\overset{\eqref{eq:22-1B}}
\le C\int_0^t\bkA{t-s}^{-9/8} \bke{n^2 |\dot u|
+ n |x \dot v|}\, ds
\\
&\overset{\eqref{uvdot.est}} \le C \int_0^t
\bkA{t-s}^{-9/8} \bke{n^2 \norm{G}_{L^{4/3}} + n^4 |x|}\, ds
\\
&\overset{\eqref{G.est}}
\le C \int_0^t \bkA{t-s}^{-9/8}
\bke{ n^4 |x| + C(C_2)n^4 \norm{\xi_0}_Y \bkA{s}^{-9/8}}\, ds
\\
& \le C_{2,6}n^4 |x(t)| + C (C_2)n^4 \norm{\xi_0}_Y \bkA{t}^{-9/8} ~,
\end{align*}

{\bf 4}. $\xithree_4+\xithree_5$. We write
$\xithree_4+\xithree_5=\int_0 ^{t} e^{-iH_0(t-s)} \Pc G_{\xi,5}(s)
\, d s$, where $G_{\xi,5}(s):=(G_\xi-i^{-1}G_3)(s)$. By Lemma
\ref{th:3-2} and Lemma \ref{th:2-2}, we have for  $t>1$,
\begin{align*}
&\norm{(\xithree_4+\xithree_5)(t)}_{L^2 \loc}
\\
&\le  \int_0 ^{t-1} \norm{e^{-iH_0(t-s)} \Pc G_{\xi,5}(s) }_{L^8}
\, d s + \int_{t-1} ^{t} \norm{e^{-iH_0(t-s)} \Pc
G_{\xi,5}(s)}_{L^4} \, d s
\\
&\le C \int_0 ^{t-1} C \length{t-s}^{-9/8}
\norm{G_{\xi,5}(s)}_{8/7} \, d s + \int_{t-1} ^{t} C
\length{t-s}^{-3/4} \norm{G_{\xi,5}(s)}_{4/3} \, d s
 \\
&\le C \bke{  \int_0 ^{t-1}  \length{t-s}^{-9/8}
+\int_{t-1} ^{t} \length{t-s}^{-3/4} } \bke{
(C_4 n^{15/4} |x(s)|+ C(C_2) n^2\norm{\xi_0} \bkA{s}^{-9/8}} \, d s
\\
&\le C_{2,7}\n^{15/4}\, |x(t)|
+ C(C_2) n^2 \norm{\xi_0}_Y \bkA{t}^{-9/8}~,
\end{align*}
for some explicit constant $C_{2,7}$.
If $t<1$, we can bound the $L^2 \loc$-norm by the $L^4$-norm.
Hence the last estimate for $t<1$ follows from the estimate
in Step 1.

We have obtained estimates on $\xi$
involving explicit constants $C_{2,1},\ldots, C_{2,7}$
and $C(C_2)$.  We now define the constant $C_2$
in \eqref{xi.est} to be:
\begin{equation}\label{eq:C2}
 C_2 \equiv C_{2,1}\, +\cdots + \, C_{2,7} ~.
\end{equation}
Since all terms involving  $C(C_2)$ have some extra $n$ factor,
$\xi$ satisfies the
estimates in  \eqref{xi.est} provided that $n$ is sufficiently
small.


Summarizing, we have proved the following lemma:

\begin{lemma}
If $n$ is sufficiently small, there is an explicit constant $C_2$
such that, if \eqref{A:pf} holds in $[0, t]$  for some  $t\le C_3 n^{-4}$,
then the estimates \eqref{xi.est} in Proposition \ref{th:3-1} also
hold in $[0,t]$.
\end{lemma}

\myremark
In the proof, we only used
\eqref{eq:3-2}, \eqref{A:pf},
Lemma \ref{th:2-2} and Lemma \ref{th:3-2}.
The information we need on the size of bound states is
in \eqref{eq:3-2} and the first estimate of \eqref{A:pf}.
Since
\eqref{eq:3-2} is always true, we only need to ensure that
the first estimate in \eqref{A:pf} holds.


\subsection{Normal form for equations of bound states}

We now compute the Poincar\'e normal form for the
bound states.  This normal form will be used to estimate
the bound states components $x$ and $y$  in next subsection.

Recall that we write
\[
x(t)= e^{-ie_0t} u(t) , \qquad y(t)= e^{-ie_1t} v(t)
\]
and the equations \eqref{u.eq} and \eqref{v.eq} for $u$ and
$v$,
\[
\dot u = -i e^{ie_0t}  \bke{ \phi_0, \, G + \Lambda_\pi } ,\qquad
\dot v = -i e^{ie_1 t} \bkt{(E_1(m)-e_1)y + (\phi_1 , G)}.
\]
Using the decompositions \eqref{La.dec} for $\Lambda_\pi$ and
\eqref{E.dec} for $E_1(m)$, we can decompose the equations for $u$
and $v$ according to  orders in $n$:
\begin{align}
\dot u &=-i e^{ie_0t} (\phi_0, G_3) -i e^{ie_0t}
(\phi_0,G_5+\Lambda_{\pi,5}) -i e^{ie_0t}
(\phi_0,G_7+\Lambda_{\pi,7}) \nonumber
\\
&\equiv R_{u,3} + R_{u,5}+ R_{u,7} ~,\label{u.eq1}
\\
\dot v&=-i e^{ie_1 t} [(\phi_1, G_3) + E_{1,2}|y|^2 y ] -i e^{ie_1
t} [(\phi_1, G_5) +  E_{1,4} |y|^4 y]\nonumber\\
&\quad
-i e^{ie_1 t} [(\phi_1, G_7) + E_1^{(6)}(|y|) \,y] \nonumber\\
&\equiv R_{v,3} + R_{v,5}+ R_{v,7}\label{v.eq1} ~.
\end{align}
Using \eqref{La5.est}, \eqref{La7.est} and Lemma \ref{th:3-2},
which assume \eqref{A:pf}, we
have
\begin{align}
|R_{u,5}| &\leC \norm{G_5}_{L^1\loc} + \norm{\Lambda_{\pi,5}}_{L^1}
\leC n^4 |x| + n^2 \norm{\xi_0}_Y \bkA{t}^{-9/8}
\label{Ru5.est}
\\
|R_{u,7}| &\leC \norm{G_7}_{L^1\loc} + \norm{\Lambda_{\pi,7}}_{L^1}
\leC n^6 |x| + n^2 \norm{\xi_0}_Y \bkA{t}^{-9/8}
\label{Ru7.est}\\
|R_{v,5}| &\leC \norm{G_5}_{L^1\loc} + |y|^5
\leC n^5 + n^2 \norm{\xi_0}_Y \bkA{t}^{-9/8}
\label{Rv5.est}\\
|R_{v,7}| &\leC \norm{G_7}_{L^1\loc} + |y|^7 \leC n^7
+ n^2 \norm{\xi_0}_Y \bkA{t}^{-9/8}
\label{Rv7.est}
\end{align}

We shall first integrate $R_{u,3}$ and $R_{v,3}$ in step 1, and
then integrate $R_{u,5}$ and $R_{v,5}$ in step 2.

 \noindent{\bf Step 1} \quad
Integration of terms of order $n^3$

In the equation of $u$, \eqref{u.eq1}, the terms of order $n^3$
are contained in $R_{u,3}=-i e^{ie_0t} (\phi_0, G_3)$.
The resonant terms from
$G_{3}$ are $|y|^2 x$ and $|x|^2x$, whose phase
cancels the factor $e^{ie_0t}$. The other four terms of order $n^3$
in $G_3$  have
different frequencies and can be exploited using integration by parts.
By \eqref{G3.def} we have
\begin{align*}
\dot u &=-i e^{ie_0t}  (\phi_0, G_3) + R_{u,5} + R_{u,7}
\\
&= c_1 \, |u|^2 u +c_2 \, |v|^2  u + \frac d{dt} (u_1^-) + g_{u,1}
+ R_{u,5} + R_{u,7}
\end{align*}
where
\begin{equation*}
 c_1=- i \la (\phi_0^2 ,\phi_0^2) , \qquad
 c_2=- i 2\la (\phi_0^2 ,\phi_1^2)
\end{equation*}
\[
u_1^- = - \bke{ \la \phi_0, \frac { e^{i(2e_0-2e_1)t} v^2 \bar u
}{2e_0-2e_1}  \phi_0 \phi_1^2 + \frac {e^{i(e_0-e_1)t} 2|u|^2 v}
{e_0-e_1}\phi_0 ^2\phi_1
 + \frac {e^{i(-e_0+e_1)t} u^2 \bar v} {-e_0+e_1} \phi_0^2
\phi_1 }
\]
and
\begin{equation*}
g_{u,1} = \Bigg( \la \phi_0, \frac { e^{i(2e_0-2e_1)t} \frac
d{dt}(v^2 \bar u) }{2e_0-2e_1}  \phi_0 \phi_1^2 + \frac
{e^{i(e_0-e_1)t} \frac d{dt}(2|u|^2 v)} {e_0-e_1}\phi_0 ^2\phi_1
 + \frac {e^{i(-e_0+e_1)t} \frac
d{dt}(u^2 \bar v)} {-e_0+e_1} \phi_0^2 \phi_1  \Bigg).
\end{equation*}

In the equation of $v$, \eqref{v.eq1},
the terms of order $n^3$ are in
$R_{v,3}= -i e^{ie_1 t} [(\phi_1, G_3) + E_{1,2}|y|^2 y ]$.
There is only one resonant term in $G_3$, namely,  $|x|^2 y$.
Another resonant term of order $n^3$ is from the  term
$E_{1,2}|y|^2 y$. The other
four terms of order $n^3$  in $G_3$ have different frequencies and can be
integrated. We thus  have
\begin{align*}
\dot v &=  -i e^{ie_1 t} [(\phi_1, G_3) + E_{1,2}|y|^2 y ] +R_{v,5}
+ R_{v,7}
\\
&= c_6  |u|^2  v +c_7 |v|^2v + \frac d{d t} ( v_1^-) + g_{v,1}
+R_{v,5} + R_{v,7}
\end{align*}
where
\begin{equation*}
 c_6=- i 2\la (\phi_0^2, \phi_1^2), \qquad
 c_7=- i E_{1,2}
\end{equation*}
\begin{align*}
v_1^- &= - \Bigg( \la \phi_1, \frac { e^{i(-e_1+e_0)t} v^2 \bar u
}{-e_1+e_0}  \phi_0 \phi_1^2 + \frac {e^{i(e_1-e_0)t} 2|v|^2 u}
{e_1-e_0}\phi_0 \phi_1^2
\\
&\qquad \qquad \qquad + \frac {e^{i(2e_1-2e_0)t} u^2 \bar v}
{2e_1-2e_0} \phi_0^2 \phi_1 + \frac {e^{i(e_1-e_0)t} |u|^2 u}
{e_1-e_0}\phi_0^3  \Bigg)
\end{align*}
and
\begin{align*}
g_{v,1} =  \Bigg( \la \phi_1, &\frac { e^{i(-e_1+e_0)t} \frac d{dt}(
v^2 \bar u) }{-e_1+e_0}  \phi_0 \phi_1^2 + \frac {e^{i(e_1-e_0)t} \frac
d{dt}(2|v|^2 u)} {e_1-e_0}\phi_0 \phi_1^2
\\
&\qquad + \frac {e^{i(2e_1-2e_0)t} \frac d{dt}(u^2 \bar v)}
{2e_1-2e_0} \phi_0^2 \phi_1 + \frac {e^{i(e_1-e_0)t} \frac d{dt}(|u|^2
u)} {e_1-e_0}\phi_0^3  \Bigg)
\end{align*}

We now define
\begin{equation}  
 u_1 = u - u_1^- ~, \qquad  v_1 = v -  v_1^- ~.
\end{equation}
The equations for $u_1$ and $v_1$ are
\begin{align*}
\dot u_1 &=  c_1  |u|^2 u + c_2 |v|^2  u
 + g_{u,1} +R_{u,5} + R_{u,7} \nonumber
\\
&= c_1  |u_1|^2 u_1 + c_2 |v_1|^2  u_1 + g_{u,2}
 + g_{u,1} +R_{u,5} + R_{u,7}  
\end{align*}
\begin{equation*} 
g_{u,2} = c_1 (|u|^2 u-|u_1|^2 u_1) + c_2 (|v|^2  u-|v_1|^2  u_1)
\end{equation*}
and
\begin{align*}
\dot v_1 &= c_6 |u|^2  v + c_7 |v|^2 v
 + g_{v,1} +R_{v,5} + R_{v,7} \nonumber
\\
&=
 c_6  |u_1|^2  v_1  + c_7 |v_1|^2 v_1 + g_{v,2}
 + g_{v,1} +R_{v,5} + R_{v,7} 
\end{align*}
\[
g_{v,2} =  c_6 (|u|^2  v-|u_1|^2  v_1) + c_7 (|v|^2 v-|v_1|^2 v_1)
~.
\]

We have finished the integration of order $n^3$ terms. Note that
both $u_1^-$ and $v_1^-$ enter the equations of
$u_1$ and $v_1$. This is the reason we compute their normal form
together.


Observe that
\begin{equation} \label{uv1.est}
|u^-_1|\leC n^2 |u|~, \qquad |v^-_1|\leC n^2 |v| ~.
\end{equation}
We now decompose $g_{u,1}$, $g_{v,1}$, $g_{u,2}$ and $g_{v,2}$
according to their orders in $n$. We want to write them as sum of order
$n^5$ and order $n^7$ terms.
We first  claim that $g_{u,1}$ and $g_{v,1}$ are of the forms
\[
g_{u,1}= e^{ie_0t} g_{u,1,5} +  g_{u,1,7} ~, \qquad
g_{v,1}= e^{ie_1t} g_{v,1,5} +  g_{v,1,7} ~,
\]
where $g_{u,1,7}$ and $g_{v,1,7}$ are order $n^7$ terms, and
$g_{u,1,5}$ and $g_{v,1,5}$ are explicit homogeneous polynomials
of degree 5 in $x,\bar x, y, \bar y$ with
{ purely imaginary} coefficients. Moreover, every term in
$g_{u,1,5}$ has a factor $x$ or $\bar x$. For example,
the first term in  $g_{u,1}$ is
\begin{align*}
&C e^{i(2e_0-2e_1)t} \frac d{dt}(v^2 \bar u)
\\
&=C e^{i(2e_0-2e_1)t} (2 \bar u v \dot v + v^2 \bar u)
\\
&=C e^{ie_0 t} \bke{ 2 \bar x y e^{-ie_1 t} \dot v
+ y^2e^{ie_0 t} \wbar{\dot u} }
\\
&=C e^{ie_0 t} \bke{ 2 \bar x y e^{-ie_1 t} [R_{v,3}+ R_{v,5}
+ R_{v,7}] + y^2e^{ie_0 t} [ \wbar {R_{u,3}}+  \wbar{R_{u,5}}
+  \wbar{R_{u,7}} ] }
\end{align*}
where $C=(\la \phi_0 , (2e_0 - 2e_1)^{-1} \phi_0 \phi_1^2)$ is real.
Repeating this calculations for all terms in $g_{u,1}$ and
collecting terms of  order $n^5$, we obtain $g_{u,1,5}$. The rest
belongs to $g_{u,1,7}$.
There are two terms of order $n^5$  in the last expression:
the terms of the form $ 2 \bar x y R_{v,3}$ and  $y^2 \wbar
{R_{u,3}}$. By definitions of $R_{u,3}$ and $R_{v,3}$,
they are explicit
polynomials of degree 5 in $x,\bar x, y, \bar y$ with
purely imaginary coefficients.

From the estimates of \eqref{Ru5.est}--\eqref{Rv7.est}, we can
bound $g_{u,1,7}$ by
\begin{equation}
\label{guv17.est}
\begin{aligned}
|g_{u,1,7}(t)| &\leC n^6 |u| + n^4 \norm{\xi_0}_Y \bkA{t}^{-9/8}
~,
\\
|g_{v,1,7}(t)| &\leC n^6 |v| + n^4 \norm{\xi_0}_Y \bkA{t}^{-9/8}
~.
\end{aligned}
\end{equation}

Similarly, we can write  $g_{u,2}$ and $g_{v,2}$ as
\[
g_{u,2}= e^{ie_0t} g_{u,2,5} +  g_{u,2,7} ~, \qquad
g_{v,2}= e^{ie_1t} g_{v,2,5} +  g_{v,2,7} ~,
\]
where
$g_{u,1,5}$ and $g_{v,1,5}$ are explicit homogeneous polynomials
of degree 5 in $x,\bar x, y, \bar y$ with
{purely imaginary} coefficients and
$g_{u,2,7}$ and $g_{v,2,7}$ are order $n^7$ terms satisfying
\begin{equation}
\label{guv27.est}
|g_{u,2,7}(t)| \leC n^6 |u| ~,\qquad
|g_{v,2,7}(t)| \leC n^6 |v|  ~.
\end{equation}
Here we have used \eqref{uv1.est} in last estimate.
Moreover, every term in
$g_{u,2,5}$ has a factor $x$ or $\bar x$. We shall not perform
calculations and estimates in details as they are similar to the previous
step. To gain some idea, we shall do one example and show it is of
the right form. By using $u-u_1 = u_1^-$,  the first
term in $g_{u,2}$ can be written as
\begin{align*}
|u|^2 u -|u_1|^2 u_1
&= u^2 \bar u - (u - u_1^-)^2 (\bar u - \wbar{u_1^-})
= u^2 \wbar{(u_1^-)} + 2 |u|^2 u_1^- + O(u |u_1^-|^2)
\\
&= x^2 \wbar{(u_1^-)} + 2 |x|^2 u_1^- + O(u |u_1^-|^2)
\end{align*}
The first two terms, $x^2 \wbar{(u_1^-)} + 2 |x|^2 u_1^-$,
contributes to $g_{u,2,5}$.
Since $u_1^-$  equals to $e^{ie_0t}$ times a polynomial of degree
$3$ in  $x,\bar x, y, \bar y$ with real coefficients and
$c_1$ in  $g_{u,2}$ is purely imaginary, they are of
the desired form.

Summarizing, we can write
\[
g_{u,1} + g_{u,2} = e^{i e_0 t} \wt R_{u,5} + g_{u,3}
\]
\[
g_{v,1} + g_{v,2} = e^{i e_1 t} \wt R_{v,5} + g_{v,3}
\]
where $\wt R_{u,5}= g_{u,1,5} + g_{u,2,5}$ and $\wt R_{v,5}=g_{u,1,5}
+ g_{u,2,5}$ are explicit homogeneous
polynomials of degree 5 in $x, \bar x, y$ and $\bar y$
with {\bf purely imaginary} coefficients.
Moreover, every term in
$\wt R_{u,5}$ has a factor $x$ or $\bar x$.
Also, $g_{u,3}= g_{u,1,7} + g_{u,2,7}$ and $g_{v,3}=g_{v,1,5}
+ g_{v,2,5}$. From the assumption  \eqref{A:pf}, we have
\begin{equation} \label{Guv5B.est}
|\wt R_{u,5}| \leC n^4|x| \, ,\quad |\wt R_{v,5}| \leC n^5.
\end{equation}
\begin{equation} \label{guv3.est}
|g_{u,3}| \leC n^6 |x| + n^4 \norm{\xi_0}_Y \bkA{t}^{-9/8}
\, ,\quad
|g_{v,3}| \leC n^7 +  n^4 \norm{\xi_0}_Y \bkA{t}^{-9/8}~.
\end{equation}
The final  equations for $u_1$ and $v_1$ are
\begin{align}
\dot u_1 &= c_1  |u_1|^2 u_1 + c_2 |v_1|^2  u_1 +
(R_{u,5} + e^{i e_0 t} \wt R_{u,5}) +  (R_{u,7} + g_{u,3} )
\label{u1.eq}
\\
\dot v_1 &= c_6  |u_1|^2 v_1 + c_7 |v_1|^2  v_1 +
(R_{v,5} + e^{i e_1 t} \wt R_{v,5}) + ( R_{v,7} + g_{v,3} )
\label{v1.eq}
\end{align}

\bigskip

\noindent{\bf Step 2} \quad Integration of terms of order $n^5$

We  now integrate terms of order $n^5$. In
$u_1$-equation \eqref{u1.eq} we have $R_{u,5} + e^{ie_0t}\wt R_{u,5}$,
where $R_{u,5}$ is
from the decomposition of original equation \eqref{u.eq1}
and $\wt R_{u,5}$ is from the error terms $g_{u,1} + g_{u,2}$.
Similarly, terms of order $n^5$ in $v_1$-equation \eqref{v1.eq}
is $R_{v,5} + e^{ie_1t}\wt R_{v,5}$.  Observe that they are either
of the form
$x^\al y^\beta$ with $|\al|+|\beta|=5$, or of the form $xy\xi$.
Also note that there are two sources in $R_{u,5} $:
$G_5$ and $\Lambda_{\pi,5}$. Among all these terms the main term
is $G_5$.

We have already studied $\wt R_{u,5}$ and $\wt R_{v,5}$.
They are explicit homogeneous
polynomials of degree 5 in $x, \bar x, y$ and $\bar y$
with {\bf purely imaginary} coefficients.
Moreover, every term in
$\wt R_{u,5}$ has a factor $x$ or $\bar x$.

We next look at $\Lambda_\pi$. Recall \eqref{La.dec} that
$\Lambda_\pi = \Lambda_{\pi,5} + \Lambda_{\pi,7}$ and
 $\Lambda_{\pi,5}=- 2 q_3 \, (\phi_1,\, G_3 |y|^2
+ \bar G_3 y^2)$ \eqref{La5.def}.  Thus $\Lambda_{\pi,5}$
is a homogeneous polynomial in $x,\bar x, y$
and $y$ of degree $5$ with purely real functions as coefficients.
Therefore the term $-i e^{ie_0t} (\phi_0,\Lambda_{\pi,5})$ in $\dot u$
equation \eqref{u.eq1}  gives only polynomials with purely imaginary
coefficients and a phase $e^{ie_0t}$.

Recall $G_5$ is given by
\begin{align*}
 G_5 &=\la (2 y^3 \bar y \bar x   + 4 |y|^4 x ) \,
\phi_0\phi_1 q_3  + \la (2|x|^2 y^2 \bar y + x^2 y \bar y^2 ) \,
\phi_0^2 q_3
\\
& \quad +  \la (x \phi_0 + y \phi_1)^2 \wbar \xi +
2\la |(x \phi_0 + y \phi_1)|^2 \xi
\end{align*}
Recall the decomposition $\xi=\xitwo + \xithree$, where
\[
\xitwo(t)= y^2 \bar x \Phi_1 + |y|^2 x \Phi_2 + |x|^2 y \Phi_3
  + x^2 \bar y \Phi_4  + |x|^2 x \Phi_5
\]
with $\Phi_1$ the only function with nonzero imaginary part
\eqref{Phi1.def}, \eqref{Phij.def}. Let
\begin{equation}
\Phi_1=\Phi_{1,R} + i \Phi_{1,I}
\end{equation}
with both $\Phi_{1,R}$ and $ \Phi_{1,I} $ real. Denote
the real part of $\xitwo(t)$ by
\begin{equation}
\xitwo_R(t)= y^2 \bar x \Phi_{1,R} + |y|^2 x \Phi_2 + |x|^2 y
\Phi_3 + x^2 \bar y \Phi_4  + |x|^2 x \Phi_5.
\end{equation}
We can write  $\xi= y^2 \bar x i \Phi_I +\xitwo_R +\xithree$.
Thus we can further decompose $G_5$ as
\begin{equation}
G_5 =G_{5,1} + G_{5,2} + G_{5,3}
\end{equation}
where
\begin{align*}
G_{5,1} &= (x \phi_0 + y \phi_1)^2 \bar y^2 x (-i) \Phi_{1,I} +
2|(x \phi_0 + y \phi_1)|^2 y^2 \bar x i \Phi_{1,I}
\\
G_{5,2} &=\la (2 y^3 \bar y \bar x   + 4 |y|^4 x ) \, \phi_0\phi_1
q_3  + \la (2|x|^2 y^2 \bar y + x^2 y \bar y^2 ) \, \phi_0^2 q_3
\\
& \quad +  \la (x \phi_0 + y \phi_1)^2 \wbar {\xitwo_R} + 2\la |(x
\phi_0 + y \phi_1)|^2 \xitwo_R
\\
G_{5,3} &=\la (x \phi_0 + y \phi_1)^2 \wbar \xithree + 2\la |(x
\phi_0 + y \phi_1)|^2 \xithree
\end{align*}

The term $G_{5,3}$ will be shown to be smaller than
$G_{5,1}$ and $G_{5,2}$. Although $G_{5,1}$
and $G_{5,2}$ are of the same size,  $G_{5,2}$ consists
of monomials in $x$, $\bar x$, $y$ and $\bar y$ with {\it real}
functions as coefficients, while $G_{5,1}$ with purely imaginary
coefficients. The reason that $G_{5,1}$ has purely imaginary
coefficients is due to the resonance of some linear
combination of eigenvalues with the continuum spectrum of $H_0$
appearing in the form $(H_0  - 0i - 2e_1 + e_0)^{-1}$.

The only resonant
term in $u$-equation from $G_{5,1}$ is $|y|^4x$ (from $y^2 \bar \xi$):
\[
  -i e^{i e_0 t} (\phi_0,(y \phi_1)^2 \bar y^2 x (-i) \Phi_{1,i})
  = - (\phi_0 \phi_1^2,\Phi_{1,i}) |v|^4 u ~,
\]
and the only resonant term in $v$-equation from $G_{5,1}$ is
$|x|^2|y|^2y$ (from $x \bar y \xi$):
\[
-i e^{i e_1 t} (\phi_1, 2 (x\phi_0)( \bar y \phi_1) y^2 \bar x i
\Phi_{1,i}) = 2 (\phi_0 \phi_1^2,\Phi_{1,i}) |u|^2 |v|^2 v ~.
\]
Note their coefficients only differ by a factor $-2$. We recall
\begin{equation}
\gamma_0 =-(\phi_0 \phi_1^2,\Phi_{1,i}) = -\Im \bke{ \la \phi_0
\phi_1^2 \, , \, \frac {-\la} {H_0  - 0i - 2e_1 + e_0} \, \Pc
\phi_0 \phi_1^2 } >0
\end{equation}
Together with the definitions of $R_{u,5}$ and
$ R_{v,5}$ in \eqref{u.eq1},  we can rewrite
\begin{align*}
&R_{u,5} + e^{i e_0 t} \wt R_{u,5} = e^{i e_0 t}
\bkt{\wt R_{u,5} -i (\phi_0,\, G_{5,1}+G_{5,2} +\Lambda_{\pi,5}) }
-i  e^{i e_0 t} (\phi_0,\,  G_{5,3})
\\
& R_{v,5} + e^{i e_1 t} \wt R_{v,5} = e^{i e_1 t}
\bkt{\wt R_{v,5} -i (\phi_1,\, G_{5,1}+G_{5,2} +E_{1,4}|y|^4y)}
-i  e^{i e_1 t} (\phi_1,\,  G_{5,3})
\end{align*}

As in Step 1, we  now integrate by parts the  non-resonant terms
inside the square brackets. The resonant terms can't be integrated and
we shall only collect them. This procedure is the same as in Step 1
and we only summarize the conclusion: there exists
constants $c_3, c_4, c_5, c_8, c_9,c_{10}$,
$u_2^-=O(u^5
+ u v^4)$ and $v_2^-=O(u^5 + u v^4)$ two homogeneous
polynomials in $u$ and $v$ of  degree $5$,
and $g_{u,4}$ and
$g_{v,4}$ the integration remainders such that
\begin{align*}
&R_{u,5} + e^{i e_0 t} \wt R_{u,5}
 = \bke{c_3 |u|^4 + c_4 |u|^2 |v|^2 + c_5|v|^4} u
\\
&\qquad   + \frac d{dt}\bke{u_2^-} + g_{u,4}
-i  e^{i e_0 t} (\phi_0,\,  G_{5,3}) ~,
\\
& R_{v,5} + e^{i e_1 t} \wt R_{v,5} =
\bke{c_8 |u|^4 + c_9 |u|^2 |v|^2 + c_{10}|v|^4} v
\\
&\qquad + \frac d{dt}\bke{v_2^-} + g_{v,4}
-i  e^{i e_1 t} (\phi_1,\,  G_{5,3})
\end{align*}
Furthermore, except $c_3$ and $c_9$, all other constants  are purely
imaginary. The real parts of $c_3$ and $c_9$ are from
$G_{5,1}$ and they are given explicitly by
\begin{equation} \label{3-46}
\Re c_3 =  \gamma_0 , \qquad  \Re c_9 =  -2\gamma_0 ~.
\end{equation}
The explicit forms of $u_2$ or $v_2$ are not important. We only
need to know their sizes.

We can now write the equations for $u$ and $v$ as
\begin{align*}
\dot u_1 &=  c_1 \, |u_1|^2 u_1  + c_2\, |v_1|^2  u_1
+ \bke{c_3 |u|^4 + c_4 |u|^2 |v|^2 + c_5|v|^4} u
\\
&\qquad   + \frac d{dt}\bke{u_2^-} + g_{u,4}
-i  e^{i e_0 t} (\phi_0,\,  G_{5,3}) + g_{u,3} + R_{u,7}
~,
\\
\dot v_1 &=  c_6 \, |u_1|^2 v_1 + c_7\, |v_1|^2  v_1
+ \bke{c_8 |u|^4 + c_9 |u|^2 |v|^2 + c_{10}|v|^4} v
\\
&\qquad + \frac d{dt}\bke{v_2^-} + g_{v,4}
-i  e^{i e_1 t} (\phi_1,\,  G_{5,3}) + g_{v,3} + R_{v,7} ~,
\end{align*}

We now define
  \begin{align}
\uu &\equiv u_1 - u_2^- =u - u_1^- - u_2^- \label{uu.def}
  \\
\vv &\equiv v_1 - v_2^- = v - v_1^- - v_2^-  \label{vv.def}
  \end{align}

We have
\begin{align*}
\dot \uu &=   c_1 \, |u_1|^2 u_1 + c_2\, |v_1|^2  u_1 + \bke{c_3
|u|^4 + c_4 |u|^2 |v|^2 + c_5|v|^4} u
\\
& \qquad
+ g_{u,4}  -i  e^{i e_0 t} (\phi_0,\,  G_{5,3}) + g_{u,3}+ R_{u,7}
\\
&=c_1 \, |\uu|^2 \uu + c_2\, |\vv|^2  \uu + \bke{c_3 |\uu|^4 + c_4
|\uu|^2 |\vv|^2 + c_5|\vv|^4} \uu + g_u
\end{align*}
and
\begin{align*}
\dot \vv &=   c_6 \, |u_1|^2 v_1 + c_7\, |v_1|^2  v_1 + \bke{c_8
|u|^4 + c_9 |u|^2 |v|^2 + c_{10}|v|^4}v
\\
&\qquad
+ g_{v,4} -i  e^{i e_1 t} (\phi_1,\,  G_{5,3}) + g_{v,3} + R_{v,7}
\\
&=c_6 \, |\uu|^2 \vv + c_7\, |\vv|^2  \vv + \bke{c_8
|\uu|^4 + c_9 |\uu|^2 |\vv|^2 + c_{10}|\vv|^4} \vv + g_v
\end{align*}
with
\begin{align}
g_u &= g_{u,4} + g_{u,5}  + g_{u,3}     + R_{u,7}
-i  e^{i e_0 t} (\phi_0,\,  G_{5,3}) \label{gu.def}
\\
g_v&= g_{v,4} +g_{v,5} + g_{v,3} + R_{v,7}
-i  e^{i e_1 t} (\phi_1,\,  G_{5,3})
\label{gv.def}
\end{align}
and
\begin{align*}
g_{u,5}&=c_1 \, (|u_1|^2 u_1-|\uu|^2 \uu)  + c_2\, (|v_1|^2
u_1-|\vv|^2  \uu) \nonumber
\\
&\quad + c_3 (|u|^4u-|\uu|^4 \uu) + c_4 (|u|^2 |v|^2 u-|\uu|^2
|\vv|^2 \uu) + c_5(|v|^4 u-|\vv|^4 \uu)
\\
g_{v,5}&=c_6 \, (|u_1|^2 v_1-|\uu|^2 \vv)  + c_7\, (|v_1|^2
v_1-|\vv|^2  \vv) \nonumber
\\
&\quad + c_8 (|u|^4v-|\uu|^4 \vv) + c_9 (|u|^2 |v|^2 v-|\uu|^2
|\vv|^2 \vv) + c_{10}(|v|^4 v-|\vv|^4 \vv) ~.
\end{align*}

Observe that the error terms $g_u$ and $g_v$ are of the form
\begin{equation*}
 g_u, g_v \sim (x^7 + x y^6) + (x^2 + y^2)\, \xithree +
(x^4 + y^4) \xi + \bke{\phi,\xi^3} + \cdots ~,
\end{equation*}
where  $\phi$ denotes some local function. For $g_v$ we should add
$|y|^7$: Since $g_v$ has a term
$-i e^{ie_1t}\, (\phi_1, E^{(6)}(|y|)y)$ from $R_{v,7}$.
This term has no factor in $x$ and is of order $|y|^7$.

Under the assumption of \eqref{A:pf}, the error terms $g_{u,4}$
and $g_{v,4}$ can
be estimated similarly as for  $g_{u,1}$ and $g_{v,1}$. Also,
$g_{u,5}$ and $g_{v,5}$
can be estimated similarly  as for $g_{u,2}$ and $g_{v,2}$.
We also have, for $j=1,2$,
\[
| e^{i e_j t} (\phi_j,\,  G_{5,3}) |
\le C n^2 \norm{\xi^{(3)}}_{L^2 \loc} \overset{\eqref{A:pf}}
\le C n^{6-1/4} |x| + C n^2  \norm{\xi_0}_Y \bkA{t}^{-9/8}~.
\]
Together with the estimates
\eqref{Ru7.est}, \eqref{Rv7.est} and \eqref{guv3.est}, we conclude
\begin{align*}
&|g_u(t)| \le C_6 \n^{6-1/4} |x(t)| + C_6 n^2 \norm{\xi_0}_Y \bkA{t}^{-9/8}
\\
&|g_v(t)| \le C_6 \n^{7-1/4}  \qquad + C_6 n^2 \norm{\xi_0}_Y \bkA{t}^{-9/8}
\end{align*}
Summarizing our effort, we have obtained the following lemma.

\begin{lemma} \label{th:3-4}

Let $\uu$ and $\vv$ be defined as in \eqref{uu.def}--\eqref{vv.def}.
They satisfy
\begin{equation} \label{eq:NF}\begin{split}
\dot \uu  &= (c_1 |\uu|^2 + c_2|\vv |^2) \uu  + (c_3 |\uu|^4 + c_4
|\uu |^2|\vv|^2 + c_{5} |\vv |^4)\uu  + g_u
\\
\dot \vv &= (c_6 |\uu|^2 + c_7 |\vv |^2) \vv + (c_8 |\uu|^4 + c_9
|\uu |^2|\vv|^2 + c_{10}|\vv |^4)\vv + g_v
\end{split}\end{equation}
All  coefficients $c_1, \cdots, c_{10}$ except $c_3$ and
$c_9$ are purely imaginary and we have \eqref{3-46}, i.e.,
\begin{equation}
\Re c_3 = \gamma_0 ~, \qquad \Re c_9 =  -2\gamma_0~,
\end{equation}
where $\gamma_0>0$ is defined in \eqref{gamma0.def}.

Moreover,
assuming \eqref{A:pf} and using the estimates \eqref{eq:3-2}
and   Lemma \ref{th:3-2}, we have
\begin{align}
&|u(t)-\uu(t)|\le C_6\,\n^2|x(t)| ~,\quad |v(t)-\vv(t)|\le C_6 \,\n^3
\label{eq:33-1} \\
&|g_u(t)| \le C_6 \n^{6-1/4} |x(t)| + C_6 n^2 \norm{\xi_0}_Y \bkA{t}^{-9/8}
\label{gu.est}\\
&|g_v(t)| \le C_6 \n^{7-1/4}  \qquad + C_6 n^2 \norm{\xi_0}_Y \bkA{t}^{-9/8}
\label{gv.est}
\end{align}
for some explicit constant $C_6$.
\end{lemma}


\subsection{Estimates for bound states}

In this subsection we will conclude the estimates for $x$ and $y$ stated
in Proposition \ref{th:3-1}.
Recall tha, t under the assumption of
\eqref{A:pf} and   \eqref{eq:3-2},  we have
proved Lemmas \ref{th:3-2}--\ref{th:3-4} which contains  estimates for
$\xi$, $g_u$ and $g_v$.

We now derive some preliminary estimates.
Let $f=2|\uu |^2$ and $g=|\vv|^2$. We have, by \eqref{eq:33-1},
\begin{equation}
\label{3-50}
\length{f(t) - 2|x|^2}\le 5 C_6 n^2 |x|^2, \qquad
\length{g(t) - |y|^2}\le 5 C_6 n^4.
\end{equation}
We also have from \eqref{eq:NF} that
\begin{align}
\dot f &=  \Re 4\bar \uu  \dot \uu  = 2\gamma_0 g^2 f + \Re 4\bar \uu
g_u  ~,\label{33-3}
\\
\dot g &=  \Re 2\bar \vv \dot \vv = -2\gamma_0 f g^2 + \Re 2\bar  \vv
g_v ~. \label{33-4}
\end{align}
By \eqref{gu.est} and \eqref{gv.est}, we have
\[
\length{\dot f+ \dot g} =
\length{\Re 4\bar \uu g_u + \Re 4\bar \vv g_v }
\le 4C_6 \n ^{8-1/4} + 4C_6 n^3 \norm{\xi_0}_Y
\bkA{t}^{-9/8} ~.
\]
Recall $n_1^2 = |x(0)|^2+ \tfrac 12 |y(0)|^2$.
By \eqref{3-50} we have  $|(f+g)(0)- 2 n_1^2| \le 10 C_6 n^4$.
Thus, for $t< C_3 n^{-4}$,
\begin{eqnarray}
|(f+g)(t)-2n_1^2| &\le& 10 C_6 n^4 + \int _0^t 4C_6 \n ^{8-1/4}
+ 4C_6 n^3 \norm{\xi_0}_Y \bkA{s}^{-9/8} ds \nonumber
\\
&\le& 5 C_3 C_6 n^{4-1/4} ~. \label{3-53}
\end{eqnarray}
We now prove Proposition \ref{th:3-1} in three steps.

\bigskip

\noindent{\bf 1. Initial layer regime}

In this period the dispersive part disperses away so much that it
becomes negligible locally. The times it  takes for this to happen
is of order $t_0 = \e_3 n^{-4}$. We first prove that
$\tfrac 12 |x_0| \le |x(t)|\le \tfrac 32 |x_0|$ for  $t \in [0,t_0]$.
The main ingredients of the proof are the norm form equation
of $x$ from the last section and the following observation.
The $\xi$ dependent term is of the form $n^2 \xi$ or of higher
orders.  Because of our assumption
$\norm{\xi_0}_Y \le \e_2 |x_0|n^{-2}$ and the decay of $\xi(t)$,
this term will not change $x(t)$ very much.
More precisely,  for $t \in [0,t_0]$, $t_0 = \e_3 n^{-4}$, we have
by \eqref{33-3}, \eqref{gu.est}, \eqref{3-50},  the assumptions
  $\norm{\xi_0}_Y \le \e_2 |x_0|n^{-2}$ and \eqref{A:pf},
\begin{eqnarray*}
|f(t) - f(0)| &\le& \int_0^{t} 4 \gamma_0 g^2 f(s) + 5 |x_0| |g_u(s)|
\, ds
\\
&\le& \bke{8  \gamma_0 n^4 + C n^{6-1/4} }  |x_0|^2 t_0 +
 10 |x_0| n^2\norm{\xi_0}_Y
\\
&\le& (10 \gamma_0+1) \e_2\,|x_0|^2 + 10 \e_3 |x_0|^2
  \le \tfrac 18 f (0),
\end{eqnarray*}
provided  $n$, $\e_2$ and $\e_3$ are sufficiently small.
By \eqref{3-50}, we have
$\length{|x(t)|^2 - |x_0|^2 } \le \tfrac 14 |x_0|^2$. Hence we
have $\tfrac 12 |x_0| \le |x(t)|\le \tfrac 32 |x_0|$ for $t \in
[0,t_0]$.

Similarly, we can show $\length{g(t) - g(0)} \le
 ((10 \gamma_0+1)  \e_2+ 10 \e_3) n^2$.
Hence we have $||y(t)|-|y_0|| \le 0.01 |y_0|$ for $t \in
[0,t_0]$ if $\e_2$ and $\e_3$ are small.
The smallness of $\e_2$ and $\e_3$ can be
guaranteed if we define
\begin{equation} \label{3-54}
  \e_2 = \frac 1{2000(\gamma_0+1)},\qquad
  \e_3 = \frac 1{2000}~.
\end{equation}

\bigskip

\noindent{\bf 2. Transition regime (i)}

In this period most mass of the disperse wave is far away and has no
effect on the local dynamics;
the ground state begins to grow exponentially until it
has the order $\n/100$. The time it takes is of order $n^{-4}$.

Define
\begin{equation} 
t_1\equiv\inf_{t \ge t_0} \bket{t:|x(t)|\ge 0.01\n} := t_1' ~.
\end{equation}
We want to show that
\begin{equation} \label{eq:33-6}
0\le t_1\le t_0 + \frac {1.01}{\;\gamma_0 \n^4\;}\;  \log \bke{\frac
{2\n} {|x_0|}} ~,
\end{equation}

Suppose \eqref{eq:33-6} fails, that is, $|x(t)|< 0.01\n$ for all
$t\le t_1'$. By \eqref{eq:33-1} and \eqref{3-50},
we have $f(t)\le 0.0004\, \n^2$
and  $g(t)\ge 0.9995 \,\n^2$ for $t\le t_1'$.
Hence
\[
\dot f(t) \ge 2 \gamma_0 (0.9995 \,\n^2)^2 f + O(\n^6)f \ge \frac
2{1.01}\gamma_0 \n^4 f ~,
\]
if $\n$ is sufficiently small. Hence
\begin{equation}\label{eq:33-7}
  f(t) \ge f(0) \exp \bket{\frac 2{1.01}\gamma_0 \n^4 t} ~,
\end{equation}
for $t\le t_1'$. We have
\[
f(t_1') \ge f(0)\exp \bket{\frac 2{1.01}\gamma_0 \n^4 \; \frac
{1.01}{\,\gamma_0 \n^4\,}\;  \log \bke{\frac \n {|x_0|}}}= f(0) \n^2
|x_0|^{-2} \ge 0.99 \n^2
\]
which is a contradiction to the assumption that $\xm(t_1')< 0.01\n$.
This shows that  $t_1$ satisfies \eqref{eq:33-6}. We also
have that \eqref{eq:33-7} holds for all $t\le t_1$, and that $f(t_1)
\ge 5 \cdot 10^{-5} \,\n^2$.

\bigskip

\noindent{\bf 3. Transition regime (ii)}

Recall the definition of $\e$ \eqref{e.def}.
Define
\begin{equation} 
t_2 \equiv\inf \bket{t:g(t ) \le (\e \n)^2} ~.
\end{equation}
We want to show that
\begin{equation} \label{eq:33-8}
t_1\le t_2\le t_1 + 10100\; (\gamma_0 \n^4 \e^2)^{-1}:= t_2' ~.
\end{equation}

Suppose the contrary, then $g(t) \ge (\e \n)^2$ for all $t\le t_2'$.
Then $\dot f >0$ and we have that $f(t) \ge 5 \cdot 10^{-5} \,\n^2$
for $t_1 \le t \le t_2'$. Hence
\[
\dot g \le  -(1.99)\, \gamma_0 f g^2 \le - 9.95 \cdot
10^{-5}\,\gamma_0 \n^2 g^2 ~.
\]
Hence
\[
g(t) \le [g(t_1)^{-1} + 9.95 \cdot 10^{-5}\,\gamma_0 \n^2
(t-t_1)]^{-1} ~, \quad (t_1 \le t \le t_2').
\]
and $ g(t_2) < (\e \n)^2$. This contradiction shows the
existence of $t_2$ satisfying \eqref{eq:33-8}.

Since $\dot g \ge - (2.01) f g^2$, similar argument shows $t_2 \ge
C\e^{-2} \n^{-4}$ if $|y(0)|> 2 \e n$.
Combining with estimate  \eqref{3-53} for $f+g$,
we have  estimates for  $f(t_2)$. From \eqref{3-50},
these estimates of $f$ and $g$ can be translated into estimates
of $x(t_2)$ and $y(t_2)$ stated in Proposition \ref{th:3-1}.

We have proved \eqref{eq:3-4},  \eqref{eq:3-5},  \eqref{eq:3-7} and
\eqref{eq:3-8} in Proposition \ref{th:3-1}, using the estimates
\eqref{eq:1-10},  \eqref{eq:3-2} and the assumption
\eqref{A:pf}. Since \eqref{A:pf} holds for $t=0$, by continuity
it holds for all $t \le t_2$.
From  Lemmas \ref{th:3-2}--\ref{th:3-4} and the above estimates,
Proposition \ref{th:3-1} is proved.


\section{Stabilization regime}

In this section we study the solution  $\psi(t)$ in the third time
regime, after the solution has become near some nonlinear ground
state. In this regime, it is natural to use the
decomposition  \eqref{psidec2} for  the solution $\psi(t)$
which emphasizes nonlinear ground states. A key issue here is to
pass the information from the  coordinates system
\eqref{psidec1} to the \eqref{psidec2}. As emphasized in the
introduction, it
is not sufficient to use only the estimates of $\psi$ at $t=t_2$.
We will also use the explicit form of the main terms in
the  dispersive part of
$\psi(t)$ to ensure that they do
not come back to affect the local dynamics, i.e., the part of the
wave represented by these terms is
``out-going'.
The set-up and proof here are similar to those in \cite{TY}
except the big terms of the dispersive part just mentioned.
We shall first  show that certain local estimates
used in \cite{TY}
are still small (see Lemma \ref{th:4-2}).

\subsection{Preliminaries}

In the time regime $t\ge t_2$, we will use the set-up in \cite{TY},
which we now briefly recall. For the proofs we refer the reader
to \cite{TY}.

For all nonlinear ground state $Q_E$ with frequency $E$,
Let $\L_E$ be the linearized operator around $Q_E$:
\begin{equation}
\label{L.def}
\L h = -i \bket{ (-\Delta + V -E + 2 \la Q_E^2)\,h + \la Q_E^2 \,\wbar h\,}
\end{equation}
With respect to $\L_E$, we can decompose
$L^2(\R^3,\Complex)$, as a real vector space, as the direct
sum of three invariant subspaces:
\begin{equation}
\label{L2.dec}
L^2(\R^3,\Complex) = S(\L_E) \oplus \eigen(\L_E) \oplus \Hc(\L_E)
\end{equation}
where $S(\L_E)$ and $\eigen(\L_E)$ are generalized eigenspaces, obtained from
perturbation of $\phi_0$ and $\phi_1$ respectively and
$\Hc(\L_E)$ corresponds
to the continuous spectrum of $\L_E$. Notice that this decomposition
is not orthogonal. Also, $ S(\L_E)= \myspan_\R(iQ_E, R_E)$, where
$R_E = \pd_E Q_E$.

For each $\psi$ sufficiently close to $Q_E$, we can decompose $\psi$ as
\begin{equation} \label{psidec2}
\psi = \bkt{Q_E + a_E R_E + \zeta_E + \eta_E}\, e^{ i \Theta_E} ~.
\end{equation}
Here $a_E, \Theta_E \in \R$ and $\zeta_E \in \eigen(\L)$ and
$\eta_E\in \Hc(\L)$.
The direction $iQ_E$ is implicitly  given
in $Q_E (e^{ i \Theta} -1)$.
Moreover, for this $\psi$ there is a unique frequency $E'$ such that
in the decomposition \eqref{psidec2} the coefficient $a$ vanishes.
In some sense it means that $Q_{E'}$ is the closest nonlinear ground
states to $\psi$.

\donothing{

Thus one can define $E(t)$ for
each $t\ge t_2$ as the unique $E$ such that the coefficient
$a$ vanishes.

For each time $T\ge t_2$, let $Q_E= Q_{E(T)}$ and $R_E=Q_{E(T)}$.
For all $t\in [t_2,T]$, we can decompose the solution $\psi(t)$
according to \eqref{psidec2} with $\L=\L_{E(T)}$:
\begin{equation} \label{eq:4-4}
\psi(t) = \bkt{Q_E + a_E(t) R_E + \zeta_E(t) +  {\eta_E} (t)}\,
e^{-E(T)t + i \Theta(t)} ~,
\end{equation}
for  $(t_2 \le t \le T)$.
Here $\zeta \in \eigen(\L_E)$, $\eta\in \Hc(\L_E)$.
}

\subsection{Estimates}

Our aim is to show that $\psi(t_2)$ satisfies the conditions
of Theorem 3 (resonance dominated solutions) in \cite{TY}.
By Proposition \ref{th:3-1} one can show that
$\psi(t_2)$ is close to a nonlinear
ground state $Q_{E_0}e^{i \Theta_0}$
in $L^2\loc$-norm, i.e.,
$\norm{Q_{E_0}}_{L^2}=n_{t_2}\le n_0$,
$\norm{\psi(t_2)-Q_{E_0}e^{i \Theta_0}}_{L^2\loc} \le \e_0 n_{t_2}$.
From the same Proposition, we have
$n_{t_2} \sim n_1 \sim n$ where $n_1$ is defined in \eqref{n1.def}
and $n$ is defined in \eqref{eq:1-10}. Thus for the  purpose of
order of magnitude, we are free to interchange $n_{t_2}$ with
$n$.  We now state the conditions
of Theorem 3 (resonance dominated solutions) in \cite{TY}:
Suppose that the initial
data $\psi (t_2)$ is decomposed as in \eqref{psidec2}
with the frequency $E=E_0$ chosen so that the coefficient $a$ vanishes.
Then the excited
state  component $\zeta$ satisfies
\begin{equation} \label{ty22}
 0 <  \norm{\zeta}
\le \e_0 \n
\end{equation}
and the dispersive part satisfies 
\begin{equation} \label{ty23}
\norm{ {\eta_E} (t_2)}_Y \le C\norm{\zeta}^2
\end{equation}
for all  $E$  close to $E_0$ with
$|E -E_0 |\le \norm{\zeta}^2$. Here the $Y$ norm is defined
in \eqref{Y.def}. We shall see that the condition \eqref{ty22}
is easy to verify by  Proposition \ref{th:3-1}. The dispersive
part, however,  is no longer localized and there is no hope
to satisfy \eqref{ty23}. In fact, even  its $L^2$-norm is not small
enough.  Recall we know from the previous
sections  that $|x(t_2)|^2 -
|x_0|^2$ is roughly one half of $|y_0|^2 - |y(t_2)|^2$. Thus, by
conservation of $L^2$-norm, the dispersive part gains the other half
of $|y_0|^2 - |y(t_2)|^2$ and $\norm{\xi(t_2)} \approx n$.

One believes on physics ground that  most mass of the
dispersive part is far away and
it has little influence to the local dynamics.
The local part of the dispersive component, on the other hand,
is generated by changes of the bound states and  is small.
Thus  the results in \cite{TY} should still hold.
This idea, however, requires to clarify the  concept of
``out-going waves''
for nonlinear equations. Instead of directly approach this
problem, we examine
 the condition \eqref{ty23} in the proof in \cite{TY}.
It is used only  to guarantee
that for all $s \ge 0$
\begin{equation} \label{eta31A}
\begin{aligned}
\norm{e^{s\L} \eta_E(t_2) }_{L^4}
&\le \tfrac 1{20}\,
\bke{(\e n)^{-2}+ \gamma_0 n^2 s} ^{-3/4+1/100} ~,
\\
\norm{e^{s\L} \eta_E(t_2) }_{L^2 \loc}
&\le C \bke{(\e n)^{-2}+ \gamma_0 n^2 s } ^{-1} ,
\end{aligned}
\end{equation}
These estimates are used to  estimate the local
contributions of $e^{s\L}  {\eta_E} (t_2) $ in (3.19) (see the following
remark) and in the
proofs for the $L^4$ and $L^2\loc$ estimates for $ {\eta_E} (t)$
in Lemmas 5.2 and
5.3.  of \cite{TY}.

{\it Remark:}
In \cite{TY}, \eqref{eta31A} is used to estimate
$\tilde \eta^{(3)}_1 (t) =e^{-iA (t-t_2)} \tilde \eta(t_2) $,
with $\tilde \eta(t_2) = e^{i \Theta(t_2)} U {\eta_E}  (t_2)$,
and $A$ being  a self-adjoint perturbation of $-\Delta+V$,
$\L = U^{-1} (-iA) U$, $U$ a bounded operator.
Since $e^{s\L} = U^{-1} e^{-isA} U$,
\[
\tilde \eta^{(3)}_1 (t)=e^{-iA (t-t_2)} \,
e^{i \Theta(t_2)} U {\eta_E}  (t_2)\
=e^{i \Theta(t_2)} U e^{(t-t_2)\L}  {\eta_E} (t_2)
\]
Thus we can choose freely to estimate either $A$ or $\L$.
Since  $e^{(t-t_2)\L}  {\eta_E} (t_2)$ is easier to estimate than
$e^{-iA (t-t_2)} \tilde \eta(t_2) $  by using
the Duhamel's expansion, we state all conditions
in terms of $\L$.

Because we only need \eqref{eta31A},
the same proof in \cite{TY} actually gives the following  stronger result.

\begin{theorem} \label{th:4-1}
Suppose that  $\psi(t_2)$ is close to a nonlinear
ground state $Q_{E_0}e^{i \Theta_0}$
in $L^2\loc$-norm, 
and suppose that in the decomposition \eqref{psidec2} with $E=E_0$
one has
$\norm{Q_{E_0}}_{L^2}=n_{t_2} \sim n \le n_0$,
$\norm{\zeta_{E_0}} \le \e_0 n$, $|a|+\norm{\eta_{E_0}}_{L^2 \loc}
\le \e_0 ^2 \n^2$.

If for all $E$ close to $E_0$ with
$|E -E_0 |\le \e_0 ^2  n^2$ the dispersive part
$ {\eta_E} (t_2)$ in the decomposition \eqref{psidec2}
satisfies the estimates \eqref{eta31A},
then the conclusion and the proof of Theorem 1
in \cite{TY} remain valid.
In particular, there is a frequency $E_\infty$ with
$|E_\infty -E_0|\le \e_0 ^2 \n^2$ and a function
$\Theta(t)= - E_\infty t + O(\log(t))$
for $t\in [t_2,\infty)$ such that
\[
 \norm{\psi(t) - Q_{E_\infty}e^{i \Theta(t)} }_{L^2 \loc}
\le C_2
\bke{(\e n)^{-2}+ \gamma_0 n^2 (t-t_2)} ^{-1/2}
\]
for some constant $C_2$.

Suppose that the initial
data $\psi (t_2)$ is decomposed as in \eqref{psidec2}
with the frequency $E$ chosen so that the coefficient $a$ vanishes.
Suppose that, in addition to the previous assumption that
\eqref{eta31A} holds for all
frequency $E$ with $|E -E_0 |\le \e_0 ^2  n^2$, the excited
state  component $\zeta$ satisfies \eqref{ty22}--\eqref{ty23}.
Then the lower bound
\[
C_1 \bke{(\e n)^{-2}+ \gamma_0 n^2 (t-t_2)} ^{-1/2}
\le \norm{\psi(t) - Q_{E_\infty}e^{i \Theta(t)} }_{L^2 \loc} 
\]
holds as well.
\end{theorem}

The merit of this modification is that we do not need the initial data
to be localized. We only need to know that its dispersive part is
``outgoing'' in certain sense. Notice that the condition on the
size of the excited component $\zeta$ is a simple consequence
of the estimates \eqref{xi.est}, \eqref{eq:3-8} and
\eqref{eq:3-4}. To see this,  we first pretend that the size of
$\zeta$ is given by $y(t_2)$ and the size of the ground state
component is given by $x(t_2)$. Then   the condition \eqref{ty22}
is just a simple consequence of \eqref{eq:3-8}. Since the
difference between the decompositions \eqref{psidec1}
and \eqref{psidec2} are higher order terms,
the condition \eqref{ty22} is easy to check.
Therefore, Theorem
\ref{th:1-1} follows from  the following Lemma:

\begin{lemma} \label{th:4-2}
Let $\psi(t)$ be the solution of \eqref{Sch} in Theorem \ref{th:1-1}
and $t_2$ be the time in Proposition \ref{th:3-1}. Let $E_0=E(t_2)$ be
the unique energy such that in the decomposition \eqref{psidec2} the
coefficient $a$ vanishes.
Then for all $E$ close to $E_0$, i.e.,  $|E-E_0|\le C \e_0^2 n^2$,
we have for all $t \ge t_2$
\begin{equation} \label{eta31B}
\norm{e^{(t-t_2)\L}  {\eta_E} (t_2) }_{L^4} \le n 
(n^2 t)^{-3/4+1/100} ~, \quad \norm{e^{(t-t_2)\L}  {\eta_E} (t_2)}_{L^2
\loc} \le  n 
(n^2 t) ^{-1} ~.
\end{equation}
\end{lemma}

Notice that, since $t_2 \sim \e^{-2} n^{-4}$ by Proposition \ref{th:3-1},
we have
$(\e n)^{-2}+ \gamma_0 n^2 (t-t_2) \sim n^{2} t$
for all $t\ge t_2$, no matter $t>2 t_2$ or $t < 2 t_2$.
Hence \eqref{eta31B} implies \eqref{eta31A} with a big margin

\noindent{\bf Proof of Lemma \ref{th:4-2}}

We have the two decompositions at  $t=t_2$
\begin{equation} 
\begin{split}
\psi(t) & = x(t)\phi_0 +Q_1(y(t))+\xi(t)
\\
& = \bkt{Q_E + a_E(t) R_E +
\zeta_E(t)+ {\eta_E} (t)}\, e^{i\Theta_E(t) }
\end{split}
\end{equation}
Since $E$ will be fixed for the rest of this proof, we shall drop
all subscripts $E$.
Hence
\begin{equation} \label{eq:4-12a}
\zeta (t)+ {\eta } (t) =\bkt{x(t)\phi_0 \, e^{-i \Theta (t)} -Q_{T}}
+\bkt{ Q_1(y(t))+\xi(t)} \, e^{-i \Theta(t)} -a_E(t) R_E~.
\end{equation}
Thus we have
\[
 {\eta } (t_2)=\PcL \bket{\bkt{x(t_2)\phi_0 \, e^{-i \Theta (t_2)} -Q_{T}}
+ \bkt{ Q_1(y(t_2))+\xi(t_2)} \, e^{-i \Theta(t_2)} }
 =\eta_{0,1} + \eta_{0,2} ~,
\]
where
\begin{align*}
\eta_{0,1}&=\PcL  \bket{\xi(t_2)e^{-i \Theta(t_2)} } ~,
\\
\eta_{0,2}&=\PcL \bket{ \bke{x(t_2)\phi_0\,e^{-i \Theta(t_2)}  - Q_E}
+Q_1(y(t_2))\, e^{-i \Theta(t_2)} } ~.
\end{align*}
Note that $\eta_{0,2}$ is a local $H^1$ function and is bounded by
$O(\n^3)$, i.e., $\norm{\eta_{0,2}}_Y \le C n^3$. Therefore we have
\begin{equation} \label{eq:4-12}
\norm{e^{(t-t_2)\L}\eta_{0,2} }_{L^4} \le C \n^3  \; \bkA{t-t_2}^{-3/4} ~,
\qquad
\norm{e^{(t-t_2)\L}\eta_{0,2} }_{L^2 \loc} \le C \n^3  \; \bkA{t-t_2}^{-3/2} ~.
\end{equation}
Hence $e^{t\L}\eta_{0,2} $ satisfies the desired estimates
with a big margin.

We now focus on the non-local term
$\eta_{0,1}=\PcL \bket{\xi(t_2) e^{-i \Theta(t_2)} }$.
Recall $\xi(t_2)$ is bounded by $2 n$ in $L^2$ \eqref{eq:3-2}
and by $4C_2 \n^{5-1/4}$ in
$L^2 \loc$ from Proposition \ref{th:3-1}.
In particular, we have $\norm{\eta_{0,1}}_{L^2} \le C
\norm{\xi(t_2)}_{L^2} \le C\n$. Thus, by Lemmas 2.6 and 2.9 of \cite{TY},
we have
$  \norm{e^{(t-t_2)\L} \eta_{0,1}}_{L^2} \le C\n $.

For convenience of notation, we write
\[
   \L = -i H_0 + iV_1 + iV_2 \conj ~,
\]
where $V_1=2\la Q_E^2$,   $V_2=\la Q_E^2$ and  $\conj$ denotes
the conjugation operator. By Duhamel's
principle,
\begin{align*}
e^{(t-t_2)\L}\eta_{0,1} &=
\PcL  e^{(t-t_2)\L} \bket{\xi(t_2) e^{-i \Theta(t_2)} }
\\
&=\PcL e^{-i(t-t_2) H_0} \xi(t_2) e^{-i \Theta(t_2)}
\\
&\quad + \int_{t_2}^t
e^{(t-s)\L} \PcL i(V_1 + V_2 \conj) e^{-i(s-t_2)H_0}
\xi(t_2) e^{-i \Theta(t_2)} \, d s ~.
\end{align*}

We now substitute \eqref{xi.eq} with $t=t_2$, i.e.,
\begin{equation} \label{eq:4-13}
\xi(t_2) =e^{-it_2H_0} \xi_0 + \int_0^{t_2}
e^{-i(t_2-\tau)H_0} \PcH G_\xi(\tau)  \, d \tau ,
\end{equation}
into the above equation. We have
\[
e^{(t-t_2)\L}\eta_{0,1} =E_1 + E_2 + E_3 + E_4
\]
where
\begin{align*}
\Omega_1 &= \PcL e^{-it H_0} \,e^{-i \Theta(t_2)}\xi_0
 \\
\Omega_2&= \int_{t_2}^t e^{(t-s)\L} \PcL i(V_1 + V_2 \conj) e^{-is H_0}
\,e^{-i \Theta(t_2)} \xi_0 \, d s
\\
\Omega_3&=\int_0^{t_2} \PcL e^{-i(t -\tau) H_0}  \PcH \, e^{-i
\Theta(t_2)} \, G_\xi(\tau) \, d \tau
\\
\Omega_4&= \int_{t_2}^t\int_0^{t_2} e^{(t-s)\L} \PcL i(V_1 + V_2 \conj)
e^{-i(s-\tau)H_0}  \PcH \,e^{-i \Theta(t_2)} \, G_\xi(\tau) \, d
\tau \,d s
\end{align*}

The only estimates we need here from \S 3 are: (cf.\eqref{Gxi.L1est})
\begin{equation}
\norm{\xi_0}_Y \le 4n , \qquad
\norm{G_\xi(\tau)}_{L^1} \le Cn^3 , \quad (0\le \tau \le t_2),\qquad
t_2 \sim \e^{-2} \n^4 ~.
\end{equation}
From the linear estimate Lemma \ref{th:2-2} we can estimate $\Omega_1$ by
\begin{align*}
&\norm{ \Omega_1  }_{L^4} \le C t ^{-3/4}\norm{ \xi_0}_{L^{4/3}}
\\
&\norm{ \Omega_1  }_{L^2 \loc} \le C \norm{ \Omega_1  }_{L^8 } \le C t
^{-9/8} \norm{ \xi_0}_{L^{8/7}}
\end{align*}
For  $\Omega_2$, since $t_2>1$, we have
\[
\norm{ \Omega_2  }_{L^4} \le C \int_{t_2}^{t} |t-s|^{-3/4} \,
|s|^{-9/8} \norm{ \xi_0}_{L^{8/7}} \, d s \le C t^{-3/4} \norm{
\xi_0}_{L^{8/7}} ~.
\]
If $t>t_2+1$,
we bound its $L^2\loc$-norm by $L^8$ for $s\in [t_2,t-1]$ and by
$L^4$ for $s\in [t-1,t]$. Thus we have
\begin{align*}
\norm{ \Omega_2  }_{L^2 \loc} &\le C \bket{ \int_{t_2}^{t-1}
|t-s|^{-9/8}
 + \int_{t-1}^t |t-s|^{-3/4} } |s|^{-9/8}\norm{ \xi_0}_{L^{8/7}}  \, d s
\\
&\le C t^{-9/8} \norm{\xi_0}_{L^{8/7}}  ~.
\end{align*}
If $t \le t_2 + 1$, we use only $L^4$ norm
for the whole interval $[t_2,t]$ and the same estimate holds.

For  $\Omega_3$, notice that $\Omega_3 = \PcL e^{-i(t-t_2)H_0} J$ where $J$
is the integral in \eqref{eq:4-13},
\[
J=  \int_0^{t_2} e^{-i(t_2-\tau)H_0} \PcH G_\xi(\tau)  \, d \tau
= \xi(t_2) - e^{-i t_2H_0} \xi_0 ~.
\]
Since it is the difference of two $L^2$ functions of order $n$,
$\norm{J}_{L^2} \le C \n$ and $\norm{ \Omega_3 (t)}_{L^2} \le C \n$ for
all $t$.

Suppose that $t\ge 2t_2$.  Since $\norm{ G_\xi}_{L^{1}}\le C \n^3$ by
\eqref{Gxi.L1est}, we have
\[
\norm{ \Omega_3 }_{L^\infty} \le C \int_0^{t_2}
|t-\tau|^{-3/2}\norm{ G_\xi(\tau)}_{L^{1}} \, d \tau
\le C \int_0^{t_2} t^{-3/2}  \n^3 \, d \tau \le C \n^3 t_2 t^{-3/2}
\]
Interpolating, we have
\[
\norm{ \Omega_3 }_{L^4}\le C \norm{ \Omega_3 }_{L^2}^{1/2}\,
\norm{ \Omega_3 }_{L^\infty}^{1/2} \le Cn^2 t_2^{1/2} t^{-3/4} ~.
\]
Since  $L^2 \loc$-norm is bounded by $L^\infty$-norm, we conclude
that
\[
\norm{ \Omega_3 }_{L^2 \loc} \le  C \n^3 t_2 t^{-3/2} \le  C \n^3
t_2^{5/8} t^{-9/8}~.
\]

Suppose now that $t<2t_2$. From similar arguments we have
\[
\norm{ \Omega_3 }_{L^4}
\le  C\int_0^{t_2} |t-\tau|^{-3/4}  \n^3 \, d \tau
\le  C\int_0^{t_2} |t_2-\tau|^{-3/4}  \n^3 \, d \tau
= C t_2^{1/4} \n^3
\]
\begin{align*}
\norm{ \Omega_3 }_{L^2 \loc}
&\le  C\int_0^{t_2-1} \norm{\cdot}_{L^8}
  +C\int_{t_2-1}^{t_2} \norm{\cdot}_{L^4}
\\
&\le  C\int_0^{t_2-1} |t_2-\tau|^{-9/8} n^3 \,d \tau
  + C\int_{t_2-1}^{t_2} |t_2-\tau|^{-3/4}  \n^3 \, d \tau
\le  C  \n^3 + C n^3
\end{align*}

We now estimate  $\Omega_4$.  Since $\norm{ G_\xi}_{L^{1}}\le C \n^3$ by \eqref{Gxi.L1est},
we have,
\begin{align*}
\norm{ \Omega_4 }_{L^4}
&\le C\int_{t_2}^t |t-s|^{-3/4} \int_0^{t_2}
|s-\tau|^{-3/2}\norm{ G_\xi(\tau)}_{L^{1}} \, d \tau \,d s
\\
&\le C\int_{t_2}^t  |t-s|^{-3/4} t_2 s^{-3/2} \n^3 \,d s
\\
&\le C t_2 \n^3   \int_{t_2}^t  |t-s|^{-3/4} s^{-9/8}t_2^{-3/8} \,d s
\\
& \le C \n^3 t_2 ^{5/8} \int_{0}^t  |t-s|^{-3/4} \bkA{s}^{-9/8} \,d s
\\
& \le C \n^3 t_2 ^{5/8} t^{-3/4}
\end{align*}

We can bound  the $L^2\loc$-norm  by $L^8$ for $s\in [t_2,t-1]$,
and by $L^4$ for $s\in [t-1,t]$ (if $t\le t_2 + 1$, we use only $L^4$ norm
for the whole interval $[t_2,t]$). Thus we have the bound
\begin{align}
\norm{ \Omega_4 }&_{L^2 \loc} \le \bke{C\int_{t_2} ^{t-1}
|t-s|^{-9/8} + C\int_{t-1}^t |t-s|^{-3/4}} \bke{\int_0^{t_2}
 |s-\tau|^{-3/2}\norm{ G_\xi(\tau)}_{L^{1}} \, d \tau}
\,d s
\nonumber
\\
&\le  C\int_{t_2} ^{t-1} |t-s|^{-9/8} (t_2 s^{-3/2} n^3) \,d s
+ C\int_{t-1}^t |t-s|^{-3/4} (t_2 s^{-3/2} n^3) \,d s
\label{eq:4-28A}
\end{align}
The second integral in \eqref{eq:4-28A} is bounded by $C\n^3 t_2 t^{-3/2}$.
The first
integral, when $t \ge 2 t_2$, is bounded by
\begin{align}
\label{eq:4-29}
&\le C\n^3 t_2 \int_{t_2}^{t/2} |t-s|^{-9/8} s^{-3/2} \, d s
+ C\n^3 t_2 \int_{t/2}^{t-1} |t-s|^{-9/8} s^{-3/2} \, d s
\\
\nonumber
&\le  C\n^3 t_2  t^{-9/8} \int_{t_2}^{t/2} s^{-3/2} \, d s
+ C\n^3 t_2 t^{-3/2} \int_{t/2}^{t-1} |t-s|^{-9/8}  \, d s
\\
\nonumber
&\le C \n^3 t_2  t^{-9/8} t_2^{-1/2} +  C \n^3 t_2 t^{-3/2}
\end{align}
On the other hand, if $t_2+1\le t \le 2 t_2$, then the first
integral in \eqref{eq:4-28A} is bounded by
\[
C\n^3 t_2 \int_{t/2}^{t-1} |t-s|^{-9/8} s^{-3/2} \, d s
\]
which is the second integral in \eqref{eq:4-29} and is
bounded by  $C \n^3 t_2 t^{-3/2}$.
Thus, for all $t \in [t_2,T]$,
\[
\norm{ \Omega_4 }_{L^2 \loc} \le
C \n^3 t_2^{1/2}  t^{-9/8}  +  C \n^3 t_2 t^{-3/2}
\]

Summarizing, we have
\[
\norm{e^{(t-t_2)\L} \eta_{0,1}}_{L^4} \le \sum_{j=1}^4
\norm{\Omega_j}_{L^4}
\le C\bket{\norm{\xi_0}_{L^{4/3}\cap L^{8/7}}
+ n^2 t_2^{1/2} +n^3 t_2^{5/8}}t^{-3/4}
\]

\[
\norm{e^{(t-t_2)\L} \eta_{0,1}}_{L^2 \loc} \le \sum_{j=1}^4
\norm{\Omega_j}_{L^2 \loc}
\le C\bket{\norm{\xi_0}_{L^{4/3}\cap L^{8/7}} +n^3 t_2^{5/8}} t^{-9/8}
\]
Since $t_2 \le \e^{-2}n ^{-4}$, equation \eqref{eta31B} holds if
$n$ is sufficiently small. We have proved Lemma \ref{th:4-2}.

\end{document}